\documentclass[preprint]{elsarticle}
\usepackage{graphicx}
\usepackage{dcolumn}
\usepackage{bm}
\usepackage[T1]{fontenc}
\usepackage[cp852]{inputenc}
\usepackage[a4paper]{geometry}
\usepackage{float}
\usepackage{units}
\usepackage{textcomp}
\usepackage{amsmath}
\usepackage{amssymb}
\usepackage{esint}
\usepackage{amsmath}   
\usepackage[all]{xy}
\usepackage{multirow, array}
 \usepackage[titletoc]{appendix}
\usepackage{rotating}

\DeclareGraphicsExtensions{.jpg}
  
\begin{document}

\title{Diffusion behavior in Nickel-Aluminium and Aluminium-Uranium diluted alloys}

\author{Viviana P. Ramunni}
\ead{vpram@cnea.gov.ar}
\address{CONICET - Avda. Rivadavia 1917, Cdad. de Buenos Aires, C.P. 1033, Argentina.} 
\address {Departamento de Materiales, CAC-CNEA, Avda. General Paz 1499, 1650 San Mart{\'i}n, Argentina. Phone number: +54 11-6772-7298; Fax: +54 11-6772-7303} 
\thanks{This work was partially financed by CONICET - PIP 00965/2010.}
\date{\today}

\begin{abstract}
Impurity diffusion coefficients are entirely obtained from a low cost classical molecular statics technique (CMST). In particular, we show how CMST is appropriate in order to describe the impurity diffusion behavior mediated by a vacancy mechanism. In the context of the five-frequency model, CMST allows to calculate all the microscopic parameters, namely: the free energy of vacancy formation, the vacancy-solute binding energy and the involved jump frequencies, from them, we obtain the macroscopic transport magnitudes such as: correlation factor, solvent-enhancement factor, Onsager and diffusion coefficients. Specifically, we perform our calculations in f.c.c. diluted $Ni-Al$ and $Al-U$ alloys. Results for the tracer diffusion coefficients of solvent and solute species are in agreement with available experimental data for both systems. We conclude that in $Ni-Al$ and $Al-U$ systems solute atoms migrate by direct interchange with vacancies in all the temperature range where there are available experimental data. In the $Al-U$ case, a vacancy drag mechanism could occur at temperatures below $550$K.  

\end{abstract}
\maketitle

\section{Introduction}

The low enrichment of $U-$Mo alloy dispersed in an $Al$ matrix is a prototype for new experimental nuclear fuels \cite{WEB}. When these metals are brought into contact, diffusion in the $Al/U-Mo$ interface gives rise to interaction phases. Also, when subjected to temperature and neutron radiation, phase transformation from $\gamma U$ to $\alpha U$ occurs and intermetallic phases develop in the U$-$Mo$/$Al interaction zone. Fission gas pores nucleate in these new phases during service producing swelling and deteriorating the alloy properties \cite{WEB,SAV05}. An important technological goal is to delay or directly avoid undesirable phase formation by inhibiting interdiffusion of $Al$ and $U$ components. Some of these compounds are believed to be responsible for degradation of properties \cite{MIR09}. 


Housseau \textit{et al.} \cite{HOU71}, based on the effective diffusion coefficients values calculated from their experimental permeation tests, have demonstrated that these undesirable phases have not influence on the mobility of $U$ in $Al$. On the other hand, Bierlin and Green \cite{BIE55} have reported the activation energy values of $U$ mobility in $Al$, based on the maximum rate of penetration of $U$ into $Al$. 

On the other hand, Brossa \textit{et al.} \cite{BRO63}, have produced couples and triplets structures using deposition methods to study the efficient diffusion barriers that should have simultaneously, a good bonding effect and a good thermal conductivity. The practical interest of a $Ni$ barrier is shown by several publications concerning with the diffusion in the systems $Al-Ni$, $Ni-U$ and $Al-Ni-U$. The study of the $Ni-Al$ binary system was, limited to solid samples of the sandwich-type, clamped together by a titanium screw and diffusion treatments have been carried out. Results from this work \cite{BRO63}, have inspired present calculations. 

Therefore it is important to study carefully and with special attention the initial microscopic processes that originate these intermetallic phases. In order to deal with this problem we started studying numerically the static and dynamic properties of vacancies and interstitials defects in the $Al$($U$) bulk and in the neighborhood of a $(111)Al/(001)\alpha U$ interface using molecular dynamics calculations \cite{PAS11,RAM10}. Here, we review our previous works \cite{PAS11,RAM10}, performing calculation of the tracer diffusion coefficients in binary $Ni-Al$ and $Al-U$ alloys, using analytical expressions of the diffusion parameters in terms of microscopical magnitudes. 

We have summarized the theoretical tools needed to express the diffusion coefficients in terms of microscopic magnitudes such as, the jump frequencies, the free vacancy formation energy and the vacancy-solute binding energy. Then we start with non-equilibrium thermodynamics in order to relate the diffusion coefficients with the phenomenological Onsager $L$-coefficients. The microscopic kinetic theory, allows us to write the Onsager coefficients in term of the jump frequency rates \cite{ALL81a,ALL81b}, which are evaluated from the migration barriers and the phonon frequencies under the harmonic approximation. The lattice vibrations are treated within the conventional framework of Vineyard \cite{VIN57} that corresponds to the classical limit.

The jump frequencies are identified by the model developed further by Le Claire in Ref. \cite{LEC78}, known as the five-frequency model for f.c.c lattices. The method includes the jump frequency associated with the migration of the host atom in the presence of an impurity at a first nearest neighbor position. All this concepts need to be put together in order to correctly describe the diffusion mechanism. Hence, in the context of the shell approximation, we follow the technique developed by Allnatt  in Refs. \cite{ALL81a,ALL81b} to obtain the corresponding transport coefficients, which are related to the diffusion coefficients through the flux equations. 

A similar procedure for f.c.c. structures was performed by Mantina et al. \cite{MAN09,MAN08} for $Mg$, $Si$ and $Cu$ diluted in $Al$ but using density functional theory (DFT). Also, using DFT calculations for b.c.c. structures, Choudhury \textit{et al.} \cite{CHO11} have calculated the tracer self-diffusion and solute diffusion coefficients in diluted $Fe-Ni$ and $Fe-Cr$ alloys including an extensive analysis of the Onsager $L$-coefficients. 

In the present work, we do not employ DFT, instead we use a classical molecular statics technique coupled to the Monomer method \cite{RAM06}. This much less computationally expensive method allows us to compute at low cost a bunch of jump frequencies from which we can perform averages in order to obtain more accurate effective frequencies. Although we use classical methods, we have also reproduced the migration barriers for $Ni-Al$ with DFT calculations coupled to the Monomer method \cite{RAM09}. 

We proceed as follows, first of all we validate the five-frequency model using the $Ni-Al$ system as a reference case for which there is a large amount of experimental data and numerical calculations \cite{ZAC12}. Since, the $Al-U$ and $Ni-Al$ systems share the same crystallographic f.c.c. structure,  the presented description is analogous for both alloys. The full set of frequencies are evaluated employing the economic Monomer method \cite{RAM06}. The Monomer is used to compute the saddle points configurations from which we obtain the jumps frequencies defined in the five-frequency model.

For the $Ni-Al$ system case, our results of the tracer solute and self-diffusion coefficients are in good agreement with the experimental data. In this case we found that $Al$ in $Ni$, at diluted concentrations, migrates as a free specie in the full range of temperatures here considered. In the case of $Al-U$, present calculations show that both, the tracer and self-diffusion coefficients agree very well with the available experimental data in Ref. \cite{HOU71}, although a vacancy drag mechanism could occur at temperatures below 500K, while, for at high temperatures the solute $U$ migrates by direct interchange with the vacancy. 

The paper is organized as follows: In Section \ref{S1} we briefly introduce a summary of the macroscopic equations of atomic transport that are provided by non-equilibrium thermodynamics \cite{ALL03,HOW64,MUR94}. In this way analytical expressions of the intrinsic diffusion coefficients in binary alloys in terms of Onsager coefficients are presented. Section \ref{S2}, is devoted to give the way to evaluate the Onsager phenomenological coefficients following the procedure of Allnat \cite{ALL81a,ALL81b} in terms of the jumps frequencies in the context of the five-frequency model. In Section \ref{S3} we show the methodology used to evaluate the tracer diffusion coefficients for the solvent and solute atoms, as well as, the so called solvent enhancement factor. Finally, in Section \ref{S5} we present our numerical results using the theoretical procedure here summarized, which show a perfect accuracy with available experimental data, also we give an expression for the vacancy wind parameter which gives essential information about the flux of solute atoms induced by vacancy flow. The last section briefly presents some conclusions. 

\section{Theory Summary: The flux equations}
\label{S1}

Isothermal atomic diffusion in binary $A-S$ alloys can be described through a linear expression between the fluxes $\vec{J}_{k}$ and the driving forces related by the Onsager coefficients $L_{ij}$ as, 
\begin{equation}
\vec{J}_{k}=\sum_{i}^{N}L_{ki}\vec{X}_{i},
\label{eq.2}
\end{equation}
where $N$ is the number of components in the system, $\vec{J}_{k}$ describes the flux vector density of component $k$, while $\vec{X}_{k}$ is the driving force acting on component $k$. The second range tensor $L_{ij}$ is symmetric ($L_{ij}=L_{ji}$) and depends on pressure and temperature, but is independent of the driving forces $\vec{X}_{k}$. From (\ref{eq.2}) the $1^{st}$ Fick's law, which describes the atomic jump process on a macroscopic scale, can be recovered. On the other hand, for each $k$ component, the driving forces may be expressed, in absence of external force, in terms of the chemical potential $\mu_{k}$, so that \cite{ALL03}, 
\begin{equation}
\vec{X}_{k}=-T\nabla\left(\frac{\mu_{k}}{T}\right). \label{eq.3}
\end{equation}
In (\ref{eq.3}) $T$ is the absolute temperature, and the chemical potential $\mu_{k}$ is the partial derivative of the Gibbs free energy with respect to the number of atoms of specie $k$ that is,
\begin{equation}
\mu_{k}=\left(\frac{\partial G}{\partial N_{k}}\right)_{T,P,N_{j\neq k}} = \mu^{\circ}_k(T,P)+k_BT\ln(c_k\gamma_k), \label{mu}
\end{equation}
where $\gamma _k$, is the activity coefficients, which is defined in terms of the activity $a_k=\gamma_{k}c_{k}$ and $c_k$, is the molar concentration of specie $k$.

For the particular case of a binary diluted alloy $(A,S)$ with $N$ available lattice sites per unit volume, containing molar concentrations $c_{A}$ for host atoms, $c_{S}$ of solute atoms (impurities) and $c_{V}$ vacancies, the fluxes in terms of the Onsager coefficients are expressed as, 
\begin{equation}
J_{A}=-\frac{k_{B}T}{N}\left(\frac{L_{AA}}{c_{A}}-\frac{L_{AS}}{c_{S}}\right)\left(1+\frac{\partial ln\gamma_{A}}{\partial lnc_{A}}\right)\nabla c_{A},
\label{JAL}
\end{equation}
\begin{equation}
J_{S}=-\frac{k_{B}T}{N}\left(\frac{L_{SS}}{c_{S}}-\frac{L_{AS}}{c_{A}}\right)\left(1+\frac{\partial ln\gamma_{S}}{\partial lnc_{S}}\right)\nabla c_{S},
\label{JBL}
\end{equation}
and 
\begin{equation}
J_V=-(J_A+J_S).
\end{equation}
From (\ref{JAL}) and (\ref{JBL}), we define 
\begin{equation}
D_{A}=\frac{k_{B}T}{N}\left(\frac{L_{AA}}{c_{A}}-\frac{L_{AS}}{c_{S}}\right)\phi_{A},
\label{DATEQ}
\end{equation}
\begin{equation}
D_{S}=\frac{k_{B}T}{N}\left(\frac{L_{SS}}{c_{S}}-\frac{L_{SA}}{c_{A}}\right)\phi_{S}.
\label{DBTEQ}
\end{equation}
In the case of $c_A,c_S>>c_V$, the diffusion coefficient for the vacancy is given by,
\begin{equation}
D_{V}=\frac{k_{B}T}{c_{V}}\left(L_{AA}+L_{SS}+2L_{AS}\right).
\label{DVTEQ}
\end{equation}
In (\ref{DATEQ}) and (\ref{DBTEQ}), $D_{A}$ and $D_{S}$ are the intrinsic diffusion coefficients for solvent $A$ and solute $S$ respectively, while $D_{V}$ is the vacancy diffusion coefficient \cite{HEU79}. In (\ref{DATEQ}) and (\ref{DBTEQ}) the quantities $\phi_A,\, \phi_S$ are the thermodynamic factors,
\begin{equation}
\phi_{A}=\left(1+\frac{\partial ln\gamma_{A}}{\partial lnc_{A}}\right)=\phi_{S}=\left(1+\frac{\partial ln\gamma_{S}}{\partial lnc_{S}}\right)=\phi_{0}. 
\end{equation}

Murch and Qin \cite{MUR94} have shown that the standard intrinsic diffusion coefficients in (\ref{DATEQ}) and (\ref{DBTEQ}) can be expressed in terms of the tracer diffusion coefficients $D^{\star}_A$, $D^{\star}_S$ which are measurable quantities, and the collective correlation factor $f_{ij}$ ($i,j=A,S$) as:
\begin{equation}
D_{A}= D^{0}_{A}\left[f_{AA}-\frac{c_A}{c_S}f^{(A)}_{AS}\right]\phi_A = D^{\star}_{A}\left[ \frac{f_{AA}}{f_A} - \left(\frac{c_A}{c_{S}}\right)\frac{f^{(A)}_{AS}}{f_A}\right]\phi_A,
\label{DATEQ1}
\end{equation}
\begin{equation}
D_{S}= D^{0}_{S}\left[f_{SS} - \frac{c_S}{c_A}f^{(S)}_{AS}\right]\phi_B = D^{\star}_{S}\left[ \frac{f_{SS}}{f_S} - \left(\frac{c_S}{c_{A}}\right)\frac{f^{(S)}_{AS}}{f_S}\right]\phi_B.
\label{DBTEQ2}
\end{equation}
The intrinsic diffusion coefficients in (\ref{DATEQ1}) and (\ref{DBTEQ2}) are known as the modified Darken equations, where $D^{0}_i=s^2\Gamma_i/6$ ($i=A,S$) are the diffusion coefficients of atoms of specie $i$ in a complete random walk performing $\Gamma_i$ jumps of length $s$ per unit time. The collective correlation factors $f_{ij}$ are related to the $L_{ij}$ coefficients through,
\begin{equation}
f_{AA} = \frac{k_BT}{Nc_A}L_{AA}\left(\frac{1}{D^0_A}\right) \,\, ;
 \,\, f_{SS} = \frac{k_BT}{Nc_S}L_{SS}\left(\frac{1}{D^0_S}\right), \label{fii}
\end{equation}
and for the mixed terms,
\begin{equation}
f^{(A)}_{AS} = \frac{k_BT}{Nc_A}L^{(A)}_{AS}\left(\frac{1}{D^0_A}\right) \,\, ; \,\, f^{(S)}_{AS} = \frac{k_BT}{Nc_S}L^{(S)}_{AS}\left(\frac{1}{D^0_S}\right). 
\label{fij} 
\end{equation}  
The tracer correlation factors $f_A$, $f_S$ are defined as the ratios $f_A=D^{\star}_{A}/D^{0}_A$ and $f_S=D^{\star}_{S}/D^{0}_S$ respectively. The term in square brackets in the second term of equations (\ref{DATEQ1}) and (\ref{DBTEQ2}), is the vacancy wind factor $G$ \cite{MAN64}. In the next sections, we present the Onsager coefficients in terms of the atomic jump frequencies taken from Ref. \cite{ALL81a,ALL81b}.

\section{The $L$-coefficients in the shell approximation}
\label{S2} 

\begin{figure}[ht]
\begin{center}
\includegraphics[width=9.0cm]{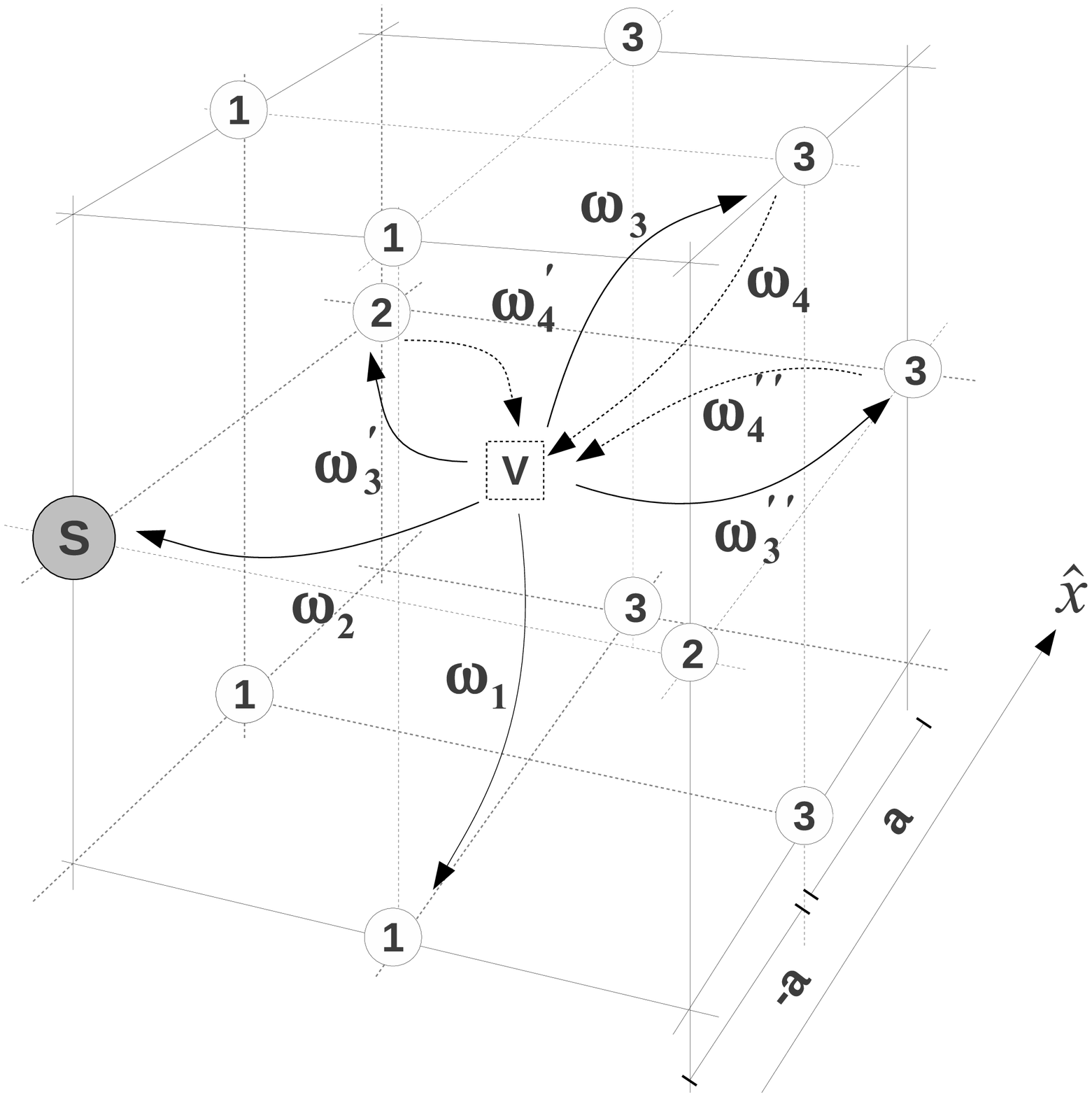}
\caption{The five-frequency model of a solute-vacancy pair in a f.c.c. lattice.}
\label{FIG1}
\end{center}
\end{figure}

\begin{figure}[h]
\begin{center}
\hspace{1.cm}\includegraphics[width=9.0cm]{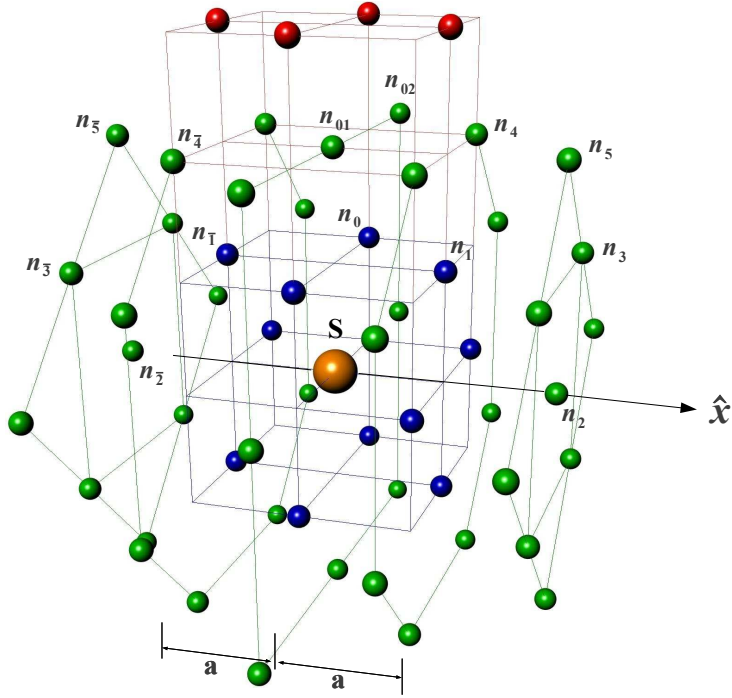}
\caption{The coordinated shell model in f.c.c. lattice (see Ref. \cite{BOC74}). The different types of symmetries shown are detailed in Table \ref{T2}. In the figure, blue bullets are the first twelve neighbors sites with the solute $\bf{S}$ at the origin. In green the 42 subsequent sites. In red, the third coordinated shell from which the vacancy never returns to the second shell. }
\label{FIG2}
\end{center}
\end{figure}

In order to understand the effect of different vacancy exchange mechanisms on solute diffusion, we adopt an effective five frequency model \`a la Le Claire \cite{LEC78} for f.c.c. lattices, assuming that the perturbation of the solute movement by a vacancy $V$, is limited to its immediate vicinity. Figure \ref{FIG1} defines the jump rates $\omega_{i}$ ($i=1,2,3,4$) considering only jumps between first neighbors.
For them, $w_{2}$ implies in the exchange between the vacancy and the solute, $w_{1}$ when the exchange between the vacancy and the solvent atom lets the vacancy as a first neighbor to the solute (positions denoted with circled 1 in figure \ref{FIG1}). The frequency of jumps such that the vacancy goes to sites that are second neighbor of the solute is denoted by $\omega_{3}$ (sites with circled 2). The model includes the jump rate $\omega_{4}$ for the inverse of $\omega_{3}$. Jumps toward sites that are third and forth neighbor of the solute are all denoted with $\omega_{3}^{\prime}$ and $\omega_{3}^{\prime\prime}$ respectively while $\omega_{4}^{\prime}$ and $\omega_{4}^{\prime\prime}$ are used for their respective inverse frequency jumps. The jump rate $\omega_{0}$ is used for vacancy jumps among sites more distant than forth neighbors of the solute atom. In this context, that enables association ($\omega_{4}$) and dissociation reactions ($\omega_{3}$), i.e the formation and break-up of pairs, the model include free solute and vacancies to the population of bounded pairs. It is assumed that a vacancy which jumps from the second to the third shell, with $\omega_0$, will never return (or returns from a random direction). As in Ref. \cite{CHO11} we express 
\begin{equation}
7\omega^{\star}_3=2\omega_3 + 4\omega^{\prime}_3 + \omega^{\prime\prime}_3,
\label{w3eff}
\end{equation}
 and 
\begin{equation}
7\omega^{\star}_4=2\omega_4 + 4\omega^{\prime}_4 + \omega^{\prime\prime}_4.
\label{w4eff}
\end{equation} 
The six symmetry types of vacancy sites that are in the first coordination shell (first neighbor with the solute) or in the second coordination shell (sites accessible from the first shell by one single vacancy jump) are shown in Figure \ref{FIG2}. Sites that are equally distant from the solute atom $S$ at the origin, and that have the same abscissa (x-coordinate in Fig.\ref{FIG2}) share the same vacancy occupation probability $n_i$, equivalently for $n_{\overline{i}}$. Table \ref{T2} resumes the sites probability with $n_{ij}$ where for $i\neq 0$ there is only one index $i$ that is given in crescent order with the distance to the solute atom $S$ in a positive abscissa, while $\overline{i}$ denote sites with negative $x$ coordinate. For the sites in the $x=0$ plane ($i=0$), the sites are denoted with two indexes as $n_{0j}$, where the second index $j$ is given in crescent order of the distance to the solute atom $S$. Table \ref{T2} denotes the number of different types of sites and the distance of them to the $x$ axis. \\
\begin{table}[h]
\begin{center}
\caption{Probability of occurrence of the vacancy at a site of the subset $n_j$.}
\label{T2}
\begin{tabular}{ccccccccc}
\hline 
 $n_{ij}$ \cite{BOC74} & \, $n_{5}$ \, &  \, $n_{4} \,$ &  \, $n_{3}$ \, &  \, $n_{2}$ \, &  \, $n_{1}$ \,  &  \, $n_{01}$ \,  &  \, $n_{02}$ \, &  \, $n_{0}$ \, \\
\hline 
\hline
\, $\#$ of sites \, & \, 4 \, &  \, 8 \, &  \, 4 \, &  \, 1 \,  &  \, 4 \,  &  \, 4 \, & \, 4 \, & \, 4 \,  
\tabularnewline
\, separation \, & \, $2a$ \, & \, $a\sqrt{2}$ \, &  \, 0 \, &  \, $a\sqrt{5}$ \, & \, $a$\, & \, $a\sqrt{2}$ \, &  \, $2a$ \, & \, $a\sqrt{2}$ \,
\tabularnewline
\hline 
\end{tabular}
\end{center}
\end{table}

\noindent The Onsager coefficients can be entirely obtained in terms of both, the free and paired specie concentrations, and the jump frequencies $\omega _i$. For the case of binary alloys the coefficients are $L_{AA}$, $L_{SS}$ and $L_{AS}$.

As was shown in Refs. \cite{ALL81a,ALL81b}, the Onsager coefficient for the solute specie can be written as
\begin{equation}
L_{SS}= L(\omega_2)\left\{1-\frac{2\omega_2}{\Omega}\right\}
\label{LBB1F}
\end{equation}
were the function $L(\omega_i)$ is,
\begin{equation}
L(\omega_i)=N\beta c_p\omega_i \frac{s^2}{6}.
\label{gW}
\end{equation}
In (\ref{gW}) $s=a/\sqrt{2}$ is the jump length, with $a$ is the lattice parameter for f.c.c. solvent $A$ and $c_p$ denotes the site fraction of solute atoms with a vacancy among their $z$ nearest-neighbor sites. $\Omega$ in (\ref{LBB1F}) is given by
\begin{equation}
\Omega=2(\omega_1+\omega_2)+7\omega^{\star}_3F.
\label{om1}
\end{equation}
Introducing $\Omega$ (\ref{om1}) in $L_{SS}$ (\ref{LBB1F}), we obtain the tracer correlation factor for the solute $f_{S}$ as,
\begin{equation}
f_{S}= \frac{2\omega_1 + 7\omega^{\star}_3F}{2(\omega_1+\omega_2)+7\omega^{\star}_3F}.
\label{eq:FF}
\end{equation}
\noindent The quantity $F$ in (\ref{eq:FF}) is a function of the ratio $y=\omega^{\star}_4 / \omega_0$ which is expressed as, 
\begin{equation}
7(1-F)=\frac{y(B_1y^{3}+ B_2y^{2}+B_3y+B_4)}{B_5y^{4}+ B_6y^{3}+B_7y^{2}+B_8y+B_9}.
\label{FMann}
\end{equation}
Table \ref{T3} shows the $B_i$ coefficients in (\ref{FMann}) calculated by Manning \cite{MAN64} and Koiwa \cite{KOI83} using respectively exact and perturbative methods. 
\begin{table}[h]
\begin{center}
\caption{Coefficients in the expression for $F$ for the five frequency model calculated by Manning \cite{MAN64} and Koiwa \cite{KOI83}.}
\label{T3}
\begin{tabular}{cccccccccc}
\hline 
\, \, & \, $B_1$ \, & \, $B_2$ \, & \, $B_3$ \, & \, $B_4$ \, & \, $B_5$ \, & \, $B_6$ \,& \, $B_7$ \, & \, $B_8$ \, & \, $B_9$ \,  
\tabularnewline
\hline 
\hline 
\,  Ref. \cite{MAN64} \, & \, 20 \, & \, 380 \, & \, 2062 \, & \, 3189 \, & \, 4 \, & \, 90 \, & \, 656 \, & \, 1861 \, & \, 1711 \, \\
\,  Ref. \cite{KOI83} \, & \, 10 \, & \, 180 \, & \, 924 \, & \, 1338 \, & \, 2 \, & \, 40 \, & \, 253 \, & \, 596 \, & \, 435 \,
\tabularnewline 
\hline 
\end{tabular}
\end{center}
\end{table}
\noindent Also following \cite{ALL81a,ALL81b}, the mixed coefficient $L_{AS}$ is,
\begin{equation}
L_{AS}=L_{SA}= 2L(\omega_2)\times\left\{3\omega^{\star}_3 -2\omega_1+7\omega^{\star}_3(1-F)(\frac{\omega_0-\omega^{\star}_4}{\omega^{\star}_4})  \right\}\frac{1}{\Omega}.
\label{LAB1F}
\end{equation}
\noindent While for the solvent, 
\begin{equation}
L_{AA}=L^{(0)}_{AA}+L^{(1)}_{AA}
\label{LAA1F}
\end{equation}
\noindent with
\begin{equation}
L^{(0)}_{AA}=L(4\omega_1+14\omega^{\star}_3)+2N\beta s^2 \omega_0 (c_V-c_p)[1-7(c_S-c_p)], 
\label{LAA2F}
\end{equation}
\noindent and
\begin{eqnarray}
L^{(1)}_{AA} = &-& 2L(3\omega^{\star}_3-2\omega_1)\left[ (3\omega^{\star}_3-2\omega_1) +
7\omega^{\star}_3(1-F)\left(\frac{\omega_0-\omega^{\star}_4}{\omega^{\star}_4}\right)
\right]\frac{1}{\Omega} \nonumber \\
& - & 2L(3\omega^{\star}_3-2\omega_1)\times \left[7\omega^{\star}_3(1-F)\left(\frac{\omega_0-\omega^{\star}_4}{\omega^{\star}_4}\right)\right]\frac{1}{\Omega} \nonumber \\
& - & 2L(3\omega^{\star}_3)\left(\frac{\omega_0-\omega^{\star}_4}{\omega^{\star}_4}\right)^2 \times \left[7(1-F)(2\omega_2+2\omega_1+7\omega^{\star}_3)\frac{1}{\Omega}\right].
\label{LAA3F}
\end{eqnarray}

For evaluating the $L$-coefficients (\ref{LBB1F}), (\ref{LAB1F}) and (\ref{LAA1F}), two parameters are needed, namely, the fraction of unbounded vacancies $c^{\prime}_V=c_V-c_p$ and the unbound solute atoms $c^{\prime}_S=c_S-c_p$. They are related with the frequency jumps through the mass action equation \cite{LEC78}, 
\begin{equation}
\frac{c_p}{c^{\prime}_Vc^{\prime}_S} = z \exp(-E_b/k_BT)=\frac{\omega^{\star}_4}{\omega^{\star}_3},
\label{cpg}
\end{equation}
where $E_b$ is the binding energy of the solute atom with a vacancy at its nearest neighbor sites. Then, if the pairs and free vacancies are in local equilibrium and the fraction of solute $c_S$ is much greater than both $c_V$ and $c_p$, we can define the equilibrium constant $K$ as,
\begin{equation}
\frac{c_p}{c_V-c_p}=zc_S\exp(-E_b/k_BT)\equiv Kc_S,
\end{equation}
and equivalently
\begin{equation}
c_p=c_V\left( \frac{Kc_S}{1+Kc_S}\right).
\label{cp0}
\end{equation}

In the next section we present the analytical expressions for the tracer diffusion coefficients $D^{\star}_A$ and $D^{\star}_S$ in terms of the jump frequencies $\omega_i$ defined in the five-frequency model through the full set of $L$-coefficients expressions in (\ref{LBB1F}-\ref{LAA3F}) and (\ref{cpg}).

\section{The tracer diffusion coefficients $D^{\star}_{A}$ and $D^{\star}_{S}$ }
\label{S3}
The diffusion model here described, is validated by the comparison of present simulations with available experimental data for the tracer diffusion coefficients $D^{\star}_{A}$ and $D^{\star}_{S}$.  \\

In the diluted limit ($c_S \rightarrow 0$) the intrinsic diffusion coefficient $D_S$ in (\ref{DBTEQ}) is identical to the tracer diffusion coefficient $D^{\star}_S$, 
\begin{equation}
D_S=D^{\star}_S(0)=\frac{k_BT}{Nc_S}L_{SS}.
\label{dif}
\end{equation}
Introducing $L_{SS}$ from (\ref{LBB1F}) in (\ref{dif}), and assuming that $c_V>> c_p \rightarrow c^{\prime}_V=c_V$ in the detailed balance equation (\ref{cpg}), we obtain an expression for the tracer solute diffusion coefficient as, 
\begin{equation} 
D^{\star}_S(0)=\frac{s^2}{6}\omega_2\left(\frac{c_p}{c_S}\right)\times \left\{\frac{2\omega_1+7\omega^{\star}_3F}{2\omega_1+2\omega_2+7\omega^{\star}_3F}\right\} \, = \, z\frac{s^2}{6}\omega_2c_V \exp(- E_b/k_BT)\times f_{S}.
\label{DBB2}
\end{equation}
where $s=a/\sqrt{2}$ and $z=12$, is the coordination number for f.c.c. lattices. In (\ref{DBB2}) the term in brackets is the solute correlation factor $f_S$. \\

On the other hand, based on Le Claire's model \cite{LEC78}, the tracer self-coefficient $D^{\star}_A(c_S)$ with a diluted concentration $c_S$ of solute atoms $S$, can be expressed in terms of the self diffusion coefficient $D^{\star}_A(0)$, of the pure $A$ matrix and the so called solvent enhancement factor $b_{A^{\star}}$ as,
\begin{equation}
D^{\star}_{A}(c_S)=D^{\star}_{A}(0)(1+b_{A^{\star}}c_S).
\label{DALC}
\end{equation}

As was shown in Ref. \cite{TUC10}, the self-diffusion coefficient $D^{\star}_{A}(0)$ in (\ref{DALC}), can be obtained from expression (\ref{DBB2}) for the tracer diffusion coefficient $S$, by replacing all the jump frequencies $\omega_i$ by $\omega_0$ and taking $E_b=0$. Hence, the self-diffusion coefficient can be written as:
\begin{equation}
D^{\star}_A(0)=z\frac{s^2}{6}\omega_0c_V f_0,
\label{DA00}
\end{equation}
where $f_0$, the correlation factor for pure f.c.c. metals, is obtained from $f_S$ in (\ref{eq:FF}) by replacing all the jump frequencies $\omega_i$ by $\omega_0$. Note that in (\ref{FMann}) if $\omega^{\star}_4/\omega_0=1$, and the $B_i$ coefficients are those in Table \ref{T3} then $7F = 5.69$ or $7F=5.15$, respectively for the Manning \cite{MAN64} or Koiwa \cite{KOI83} descriptions. Inserting the value $7F=5.69$ or $7F=5.15$ in (\ref{eq:FF}) we obtain $f_0=0.7936$ or $f_0=0.7814$, respectively. 

At thermodynamic equilibrium the vacancy concentration $c_V=c^{(0)}_V$ is given by,
\begin{equation}
c^{0}_V=\exp\left(-E^V_f /k_BT\right),
\label{cv0}
\end{equation}
where $E^V_f$ is the formation energy of the vacancy in pure $A$. The entropy terms are here set to zero, which is a simplifying approximation. So that, inserting (\ref{cv0}) in (\ref{DA00}) we get 
\begin{equation}
D^{\star}_A(0)=z\frac{s^2}{6}\omega_0f_0\exp\left( -\beta E^V_f \right).
\label{DACB}
\end{equation}

As was demonstrated by Le Claire in Ref. \cite{LEC78}, the solvent enhancement factor, $b_{A^{\star}}$ in (\ref{DALC}), depends on the properties of the solute-vacancy model. As an approximation for the five-frequency model, only valid in the context of the random alloy model \cite{ALL03}, $b_{A^{\star}}$ can be calculated directly from the Onsager phenomenological coefficients $L_{AS}$ and $L_{AA}$ in (\ref{LAB1F}) and (\ref{LAA1F}) respectively, through,
\begin{equation}
D^{\star}_A= \frac{k_BTf_0}{Nc_A}(L_{AA}+L_{AB}).
\label{SumR}
\end{equation}
Then, $b_{A^{\star}}$ is obtained by equating the expressions (\ref{DALC}) and (\ref{SumR}) for $D^{\star}_A$ hence, 
\begin{equation}
\frac{k_BT}{Nc_A}\left(L_{AA}+L_{AS}\right)=z\frac{s^2}{6} \omega_0c_V(1+b_A^{\star} \,.\, c_S).
\label{SumRbA}
\end{equation} 

Also, Belova and Murch \cite{BEL03} have address the problem of the enhancement of the solvent in diluted alloys giving an expression for $b_{A^{\star}}$ in terms of $f_0$ and the ratio $\omega_2/\omega_0$, up to third order in the solute concentration. The authors \cite{BEL03} have then obtained an excellent agreement with the theory of Moleko \textit{et al.} \cite{MOL89}.  

In more concentrated alloys the understanding of the diffusion behavior requires a significantly different approach as the one developed by Van der Ven \textit{et al.} in Refs. \cite{VEN05,VEN08}. Recently, Van der Ven \textit{et al.} \cite{VEN10}, gave another point of view of the same transport phenomena, describing a formalism to predict diffusion coefficients of substitutional alloys from first principles restricted to vacancy mediated diffusion mechanism. This approach relies on the evaluation of Kubo-Green expressions of kinetic transport coefficients using Monte Carlo simulations. 

\section{Results}
\label{S5}
 
We present our numerical results, using a classical molecular static technique (CMST) coupled to the Monomer method \cite{RAM06}, applied to $Ni-Al$ and $Al-U$ diluted alloys. In the case of the $Ni-Al$ system, for the pure elements $Ni$ and $Al$, as well as, for the cross $Ni-Al$ term, the atomic interaction are represented by EAM potentials, developed by Mishin \textit{et al.} \cite{MIS99}, where the cross term, was fitted taking into account the available first principles data. For the $Al-U$ system concerning to the pure elements, we use the potential developed by Zope and Mishin \cite{ZOP03} for $Al$, while for $U$ and the cross term we use the potentials reported in Ref. \cite{PAS12}. In this case, lattice parameters, formation energies and bulk modulus for each intermetallic compound are well reproduced. The cross potential in Ref. \cite{PAS12}, has been fitted taking into account the available first principles data \cite{ALO09}. We obtain the equilibrium positions of the atoms by relaxing the structure via the conjugate gradients technique. The lattice parameters that minimize the crystal structure energy are $a_{Ni}=3.52\, $\AA\, for $Ni$ and $a_{Al}=4.05\, $\AA\, for $Al$. For all calculations we use a christallyte of $8 \times 8 \times 8$ of 2048 atoms, with periodic boundary conditions.

Impurity and defect relaxation, includes one substitutional $Al$ atom in $Ni$ or one substitutional $U$ atom in $Al$, as well as, a single vacancy. 
Current calculations have been performed at $T=0K$. In this case, the entropic barrier is ignored. Our calculations are carried out at constant volume, and therefore the enthalpic barrier $\Delta H=\Delta U +p\Delta V$ is equal to the internal energy barrier $\Delta U$. 

In Table \ref{T4}, we present our results for the vacancy formation energy ($E^V_f$) in pure $Ni$ and $Al$ calculated as $E^V_f=E(N-1)+E_c-E(N)$, where $E(N)$ is the energy of the perfect lattice of $N$ atoms, $E(N-1)$ is the energy of the defective system, and $E_c$ the cohesion energy. The vacancy migration barrier in perfect lattice, $E^V_m$, is calculated with the Monomer method \cite{RAM06}, and the activation energy, $E_Q$, is then obtained as, $E_Q=E^V_f+E^V_m$.
\begin{table}[ht] 
\begin{center}
\caption{Energies and lattice parameters for the pure f.c.c. $Al$ and $Ni$ and $\alpha U$ lattices. The first column specifies the metal, vacancy formation energy $E^V_f(eV)$ are shown in the second column. The third column displays the migration energies $E^V_m$, calculated from the Monomer method \cite{RAM06}. In the forth column we show the lattice parameter $a_{A}$(\AA). The last column displays the activation energy $E_Q(eV)$.}
\label{T4}
\begin{tabular}{lcccccc} 
\hline 
Reference \, & \, Latt. \, & \, $E_c(eV)$ \, &  \, $E^V_f(eV)$ \, & \, $E^V_m (eV)$  \, & \, $a_{A}$(\AA) \, & $E_Q(eV)$\\
\hline 
\hline
 $\mathbf{Ni-Al}$  &&&&&& \\ 
Present work \, & \, $Ni$ \, & \, -4.45 \, & \, 1.56 \, & \, 0.98 \, & \, 3.52 \, & 2.54 \\
Voter and Chen \cite{VOT87} \, & \, $Ni$ \, & \,  -4.45 \, & \, 1.56 \, & \, 0.98 \, & \, 3.52 \, & 2.54 \\ 
Ref. \cite{MIS99} using CMST \, & \, $Ni$ \, & \, -4.45 \, & \, 1.60 \, & \, 1.29 \, & \, 3.52 \, & 2.89 \\
Ref. \cite{ZAC12} using VASP \, & \, $Ni$ \, & \, -4.45 \, & \, 1.40 \, & \, 1.28 \, & \, 3.52 \, & 2.65 \\
Experimental/ab-initio \, & \, $Ni$ \, & \, -4.45 \cite{SMI76} \, & \, 1.60 \cite{WYC78} \, & \, 1.30 \cite{MUR75} \, & \, 3.52 \cite{MUR75} \, & 2.90 \\
\,Present work \, & \, $Al$ \, & \,  -3.36  \, & \, 0.68 \, & \, 0.65 \, & \, 4.05 \, & 1.33 \\
Voter and Chen \cite{VOT87} \, & \, $Al$ \, & \,  -3.36 \, & \, 0.63 \, & \, 0.30 \, & \, 4.05 \, & 0.93 \\
Ref. \cite{MIS99} using CMST \, & \, $Al$ \, & \, -3.36 \, & \, 0.68 \, & \, 0.64 \, & \, 4.05 \, & 1.32 \\
Experimental/ab-initio \, & \, $Al$ \, & \,  -3.36 \cite{SMI76} \, & \, 0.68 \cite{WYC78} \, & \, 0.65 \cite{MUR75} \, & \, 4.05 \cite{KIT86} \, & 1.33 \\ 
\hline
 $\mathbf{Al-U}$  &&&&&& \\
Present work \, & \, $Al$ \, & \, -3.36 \, & \, 0.65 \, & \, 0.65 \, & \, 4.05 \, & 1.30 \\
Ref. \cite{ZOP03} using CMST \, & \, $Al$ \, & \, -3.36 \, & \, 0.68 \, & \, 0.63 \, & \, 4.05 \, & 1.31 \\
Present work \, & \, $\alpha U$ \, & \, -5.77 \, & \, 1.36 \, & \, 0.23 \, & \, $a_U=2.77$  \, & 1.59 \\
 \, & \, \, & \,  \, & \, \, & \, \, & \, $b_U=6.07$  \, & \\
 \, & \, \, & \,  \, & \, \, & \, \, & \, $c_U=4.94$  \, &  \\
\hline
\end{tabular}
\end{center}
\end{table} 

For the case of a diluted alloy, we consider the presence of solute vacancy complexes, $C_n=S+V_n$, in which $n=1^{st}, 2^{nd}, 3^{rd},\dots$ (see the insets in Table \ref{T5}) indicates that the vacancy is a $n-$nearest neighbors of the solute atom $S$. The binding energy between the solute and the vacancy for the complex $C_n=S+V_n$ in a matrix of $N$ atomic sites is obtained as,

\begin{equation}
E_b=\left\{ E(N-2,C_n) + E(N)\right\} - \left\{ E(N-1,V) + E(N-1,S)\right\} ,
\label{Ebnn}
\end{equation}
where $E(N-1,V)$ and  $E(N-1,S)$  are the energies of a crystallite containing ($N-1$) atoms of solvent $A$ plus one vacancy $V$, and one solute atom $S$ respectively, while $E(N-2,C_n)$ is the energy of the crystallite containing ($N-2$) atoms of $A$ plus one solute vacancy complex $C_n=S+V_n$. With the sign convention used here $E_b <0$ means attractive solute-vacancy interaction, and $E_b>0$ indicates repulsion. 

\noindent For the alloys, we calculate the migration energies $E_m$ using also the Monomer Method \cite{RAM06}, a static technique to search the potential energy surface for saddle configurations, thus providing detailed information on transition events. The Monomer computes the least local curvature of the potential energy surface using only forces. The force component along the corresponding eigenvector is then reversed (pointing ``up hill"), thus defining a pseudo force that drives the system towards saddles. Both, local curvature and configuration displacement stages are performed within independent conjugate gradients loops. The method is akin to the Dimer one from the literature \cite{HEN01}, but roughly employs half the number of force evaluations which is a great advantage in ab-initio calculations. 

\noindent Tables \ref{T5} and \ref{T6} display, respectively for $Ni-Al$ and $Al-U$, the different type of solute vacancy complexes $C_n=S+V_n$ with its binding energies $E_b$ and with the corresponding jump frequencies. Also, the same tables, depict the possibles configurations and jumps involved.

For $Ni-Al$, a weak binding energy, $E_b$, can be observed for almost all the solute-vacancy complexes, $C_n$, being attractive for $C_1$ and $C_4$ and repulsive for the rest of the pairs. The same behavior is observed in $Al-U$, although for this case, the binding energy, $E_b$, for the $C_1$ complex is strongly attractive. 

Concerning with the migration barriers, summarized in Table \ref{T5}, our results show that for $Ni-Al$, the migration barriers $E^{\leftarrow}_m$ are close to the perfect lattice value ($0.98eV$). 

\clearpage
\begin{table}[htdp] 
\begin{center}
\caption{ Jumps and frequencies in $Ni-Al$. The first column denotes $C_n=S+V_n$ where $V_n$ means that the vacancy is $n$ nearest neighbor of the solute. Binding energy $E_b$ is shown in the second column. The jumps are depicted in the third column, while the forth column describes the jump frequency $\omega_i$ and the configurations involved in each jump. Migration energies $E_m$ for direct and reversed jumps are written in the fifth and sixth column respectively.}
\label{T5}
\begin{tabular}{cccccc} 
\hline 
\, $C_n=S+V_n$ \, & \, $E_b(eV)$ &Config.($F_n$)& \, $\omega_i $ \, & \, $E^{\rightarrow}_m(eV)$ \, & \, $E^{\leftarrow}_m(eV)$ \, \\
\hline 
\, $C_1$ \, & \, -0.06 \, & 
\begin{minipage}{3.1cm} 
 \includegraphics[width=3.1cm]{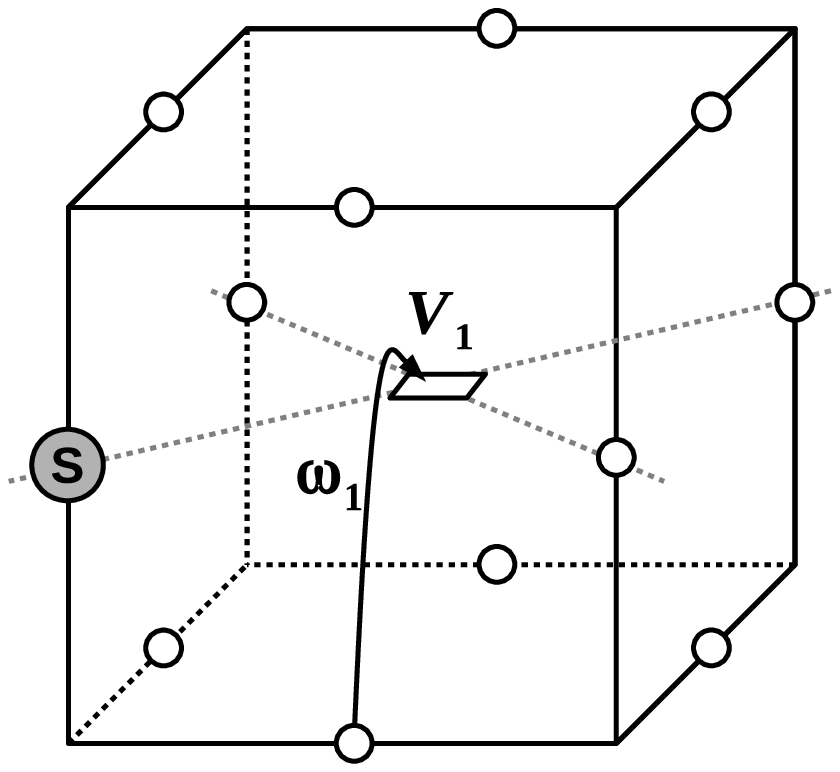} 
\end{minipage}&
\, $\xymatrix{
C_1  \ar@<0.5ex>[r]^{\omega_{1}}
& C_1 \ar@<0.5ex>[l]^{\omega_{1}} }
$ \, & \, 1.09 \, & \, 1.09 \,  \\
\, $C_{1S}$ \, & \, -0.06 \, & 
\begin{minipage}{3.1cm} 
 \includegraphics[width=3.1cm]{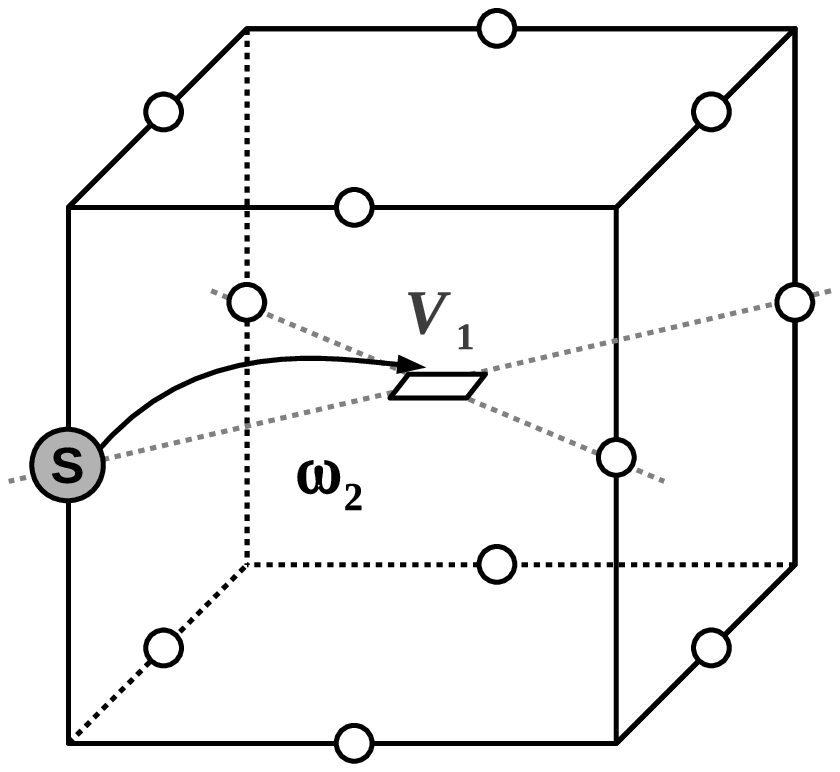} 
\end{minipage}&
\, $\xymatrix{
C_{1S}  \ar@<0.5ex>[r]^{\omega_{2}}
& C_{1S} \ar@<0.5ex>[l]^{\omega_{2}} }
$ \, & \, 0.97 \, & \, 0.97 \, \\
\, $C_2$ \, & \, 0.03 \, & 
\begin{minipage}{3.1cm} 
 \includegraphics[width=3.1cm]{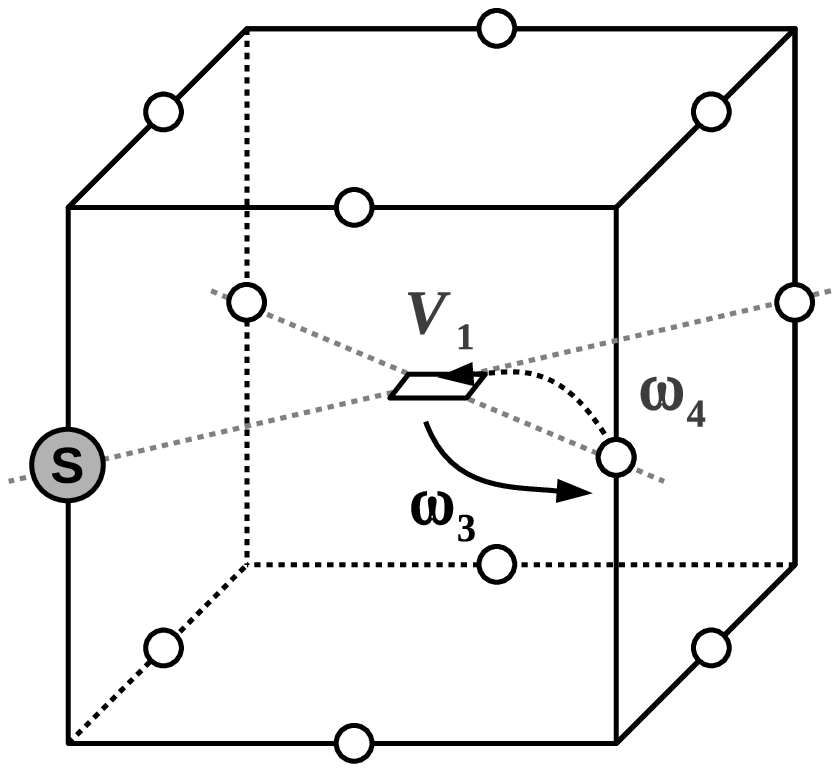} 
\end{minipage}&
\, $\xymatrix{
C_1  \ar@<0.5ex>[r]^{\omega_{3}}
& C_2 \ar@<0.5ex>[l]^{\omega_{4}} }
$ \, & \, 0.98 \, & \, 0.89 \, \\
\, $C_3$ \, & \, 0.03 \, & 
\begin{minipage}{3.1cm} 
\includegraphics[width=3.1cm]{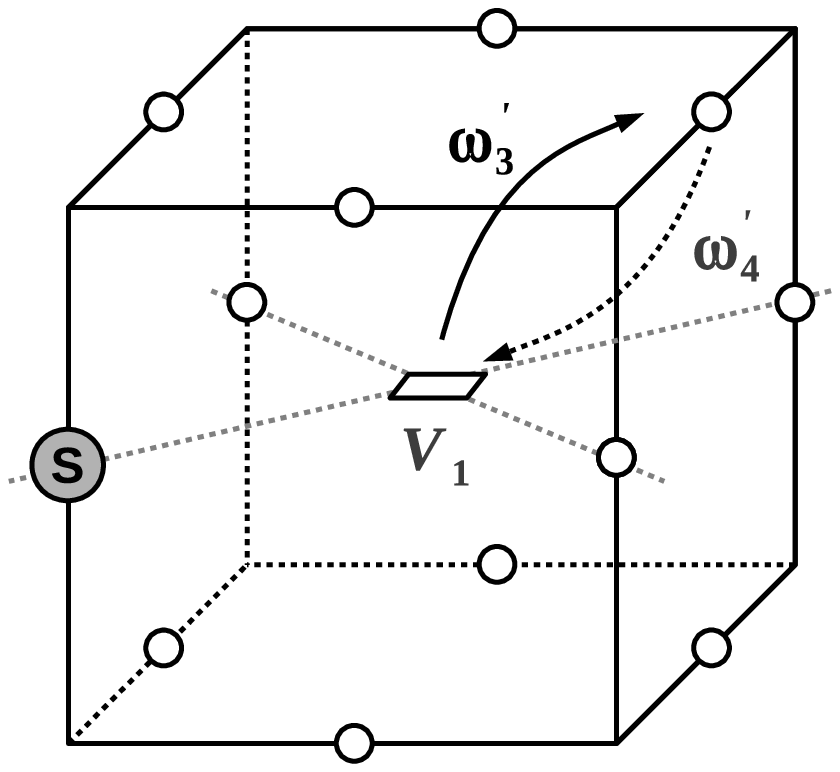} 
\end{minipage}&
\, $\xymatrix{
C_1  \ar@<0.5ex>[r]^{\omega^{\prime}_{3}}
& C_3 \ar@<0.5ex>[l]^{\omega^{\prime}_{4}} }
$ \, & \, 0.99 \, & \, 0.91 \, \\
\, $C_4$ \, & \, -0.001 \, & 
\begin{minipage}{3.1cm} 
\includegraphics[width=3.1cm]{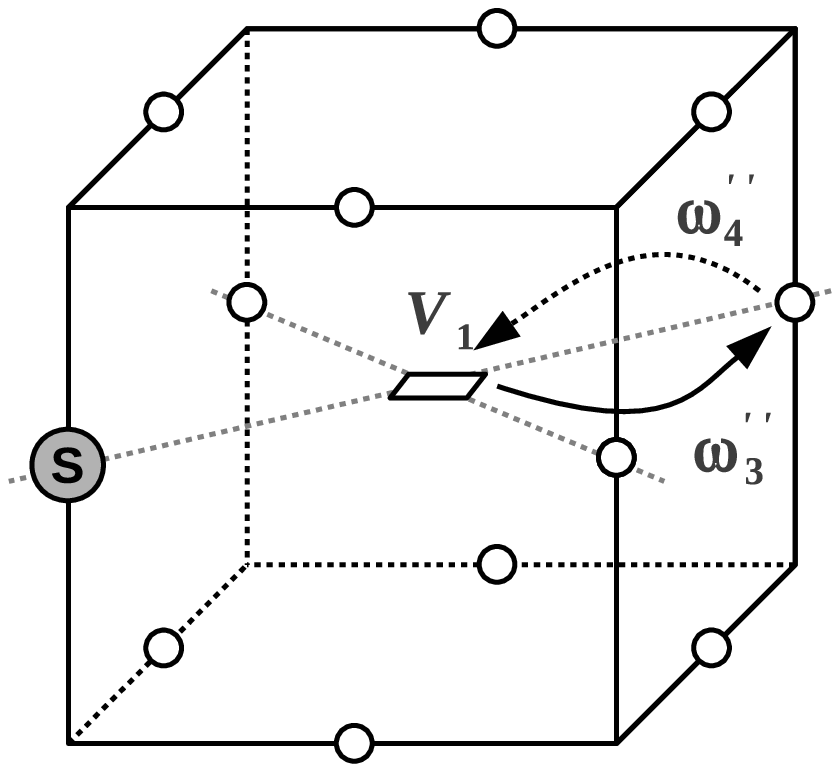} 
\end{minipage}&
\, $\xymatrix{
C_1  \ar@<0.5ex>[r]^{\omega^{\prime\prime}_{3}}
& C_4 \ar@<0.5ex>[l]^{\omega^{\prime\prime}_{4}} }$ \, & \, 0.96 \, & \, 0.90 \, \\
\, $C_5$ \, & \, $0.034$ \, & 
\begin{minipage}{3.4cm} 
 \includegraphics[width=3.4cm]{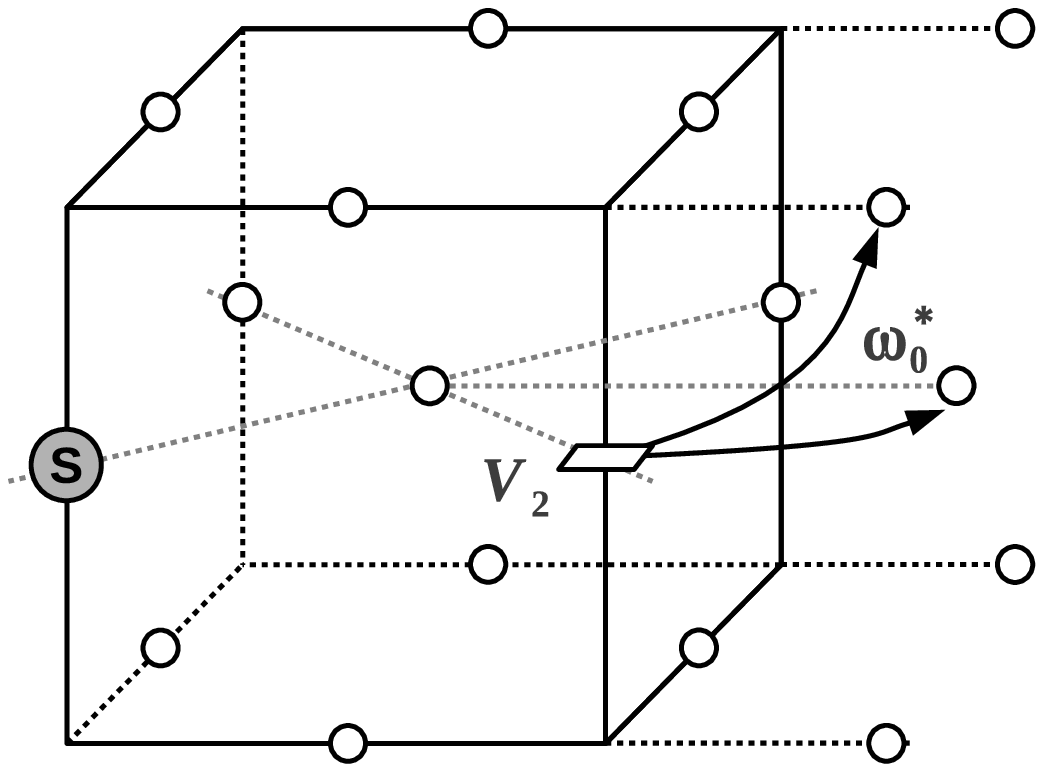} 
\end{minipage}&
\, $\xymatrix{
C_2  \ar@<0.5ex>[r]^{\omega^{\star}_{0}}
& C_5 \ar@<0.5ex>[l]^{\omega^{\star}_{0}} }$ \, & \, 0.89 \, & \, 0.98 \, \\
\, $C_6$ \, & \, $0.031$ \, & 
\begin{minipage}{3.4cm} 
\includegraphics[width=3.4cm]{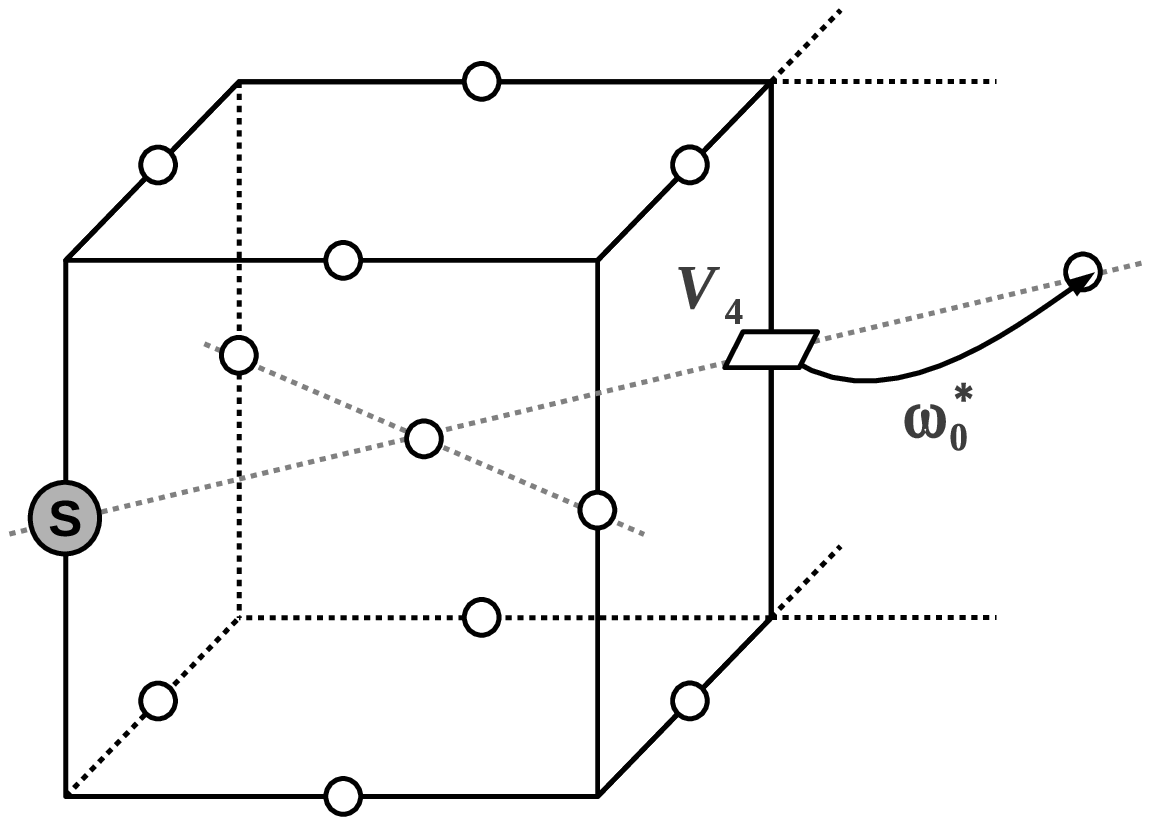} 
\end{minipage}&
\, $\xymatrix{
C_4  \ar@<0.5ex>[r]^{\omega^{\star}_{0}}
& C_6 \ar@<0.5ex>[l]^{\omega^{\star}_{0}} }$ \, & \, 0.98 \, & \, 0.98 \, \\
\, $C_7$ \, & \, $-0.001$ \, & 
\begin{minipage}{3.1cm} 
\includegraphics[width=3.1cm]{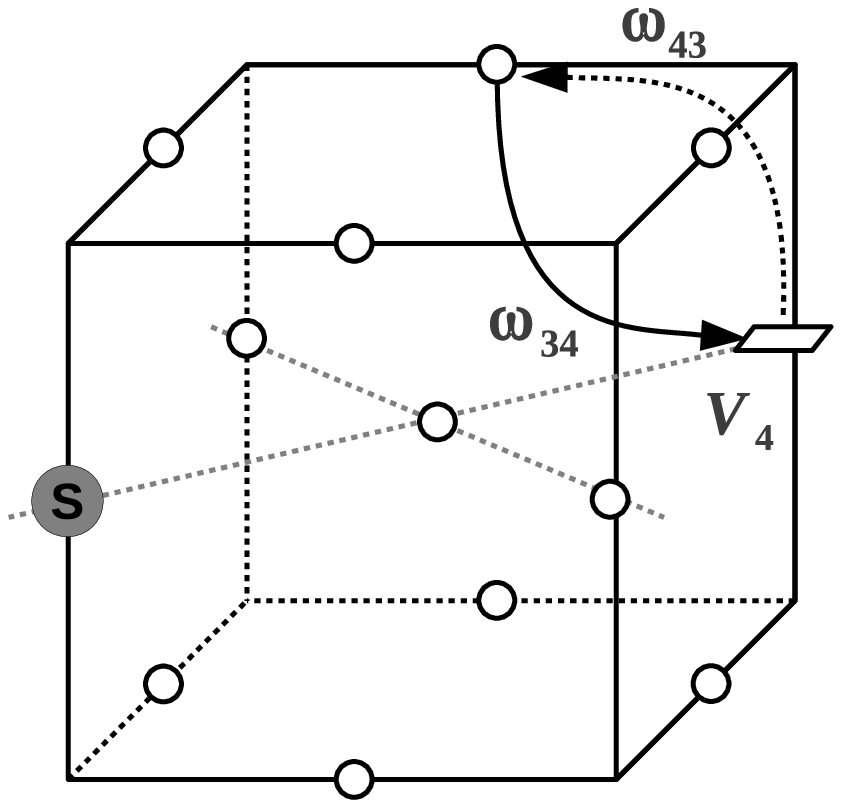} 
\end{minipage}&
\, $\xymatrix{
C_2  \ar@<0.5ex>[r]^{\omega_{43}}
& C_5 \ar@<0.5ex>[l]^{\omega_{34}} }$ \, & \, 1.01 \, & \, 0.98 \, \\
\hline
\end{tabular}
\end{center}
\end{table}

\clearpage

\begin{figure}[h]
\begin{center}
\includegraphics[width=6.0cm]{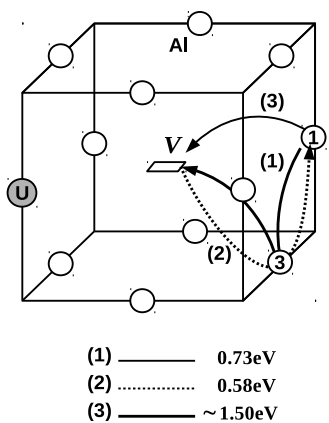}
\caption{Single jump involving two atoms in $Al-U$. In jump (1) the atom labeled 1 takes the place of atom 3, which is dragged by the atom 1 towards the vacancy $V$. Jump (2) is the reverse of jump (1). We also depict a direct jump (3) which is a high energy jump involving only the atom 1.}
\label{MJ}
\end{center}
\end{figure}

\begin{table}[htdp] 
\begin{center}
\caption{Jumps and frequencies in $Al-U$. The columns description is the same as in Table \ref{T5}.}
\label{T6}
\begin{tabular}{cccccc} 
\hline 
\, $C_n=S+V_n$ \, & \, $E_b(eV)$ &Config.($F_n$)& \, $\omega_i $ \, & \, $E^{\rightarrow}_m(eV)$ \, & \, $E^{\leftarrow}_m(eV)$ \, \\
\hline 
\, $C_1$ \, & \, -0.139 \, & 
\begin{minipage}{3.50cm} 
 \includegraphics[width=3.50cm]{FIG5.eps} 
\end{minipage}&
\, $\xymatrix{
C_1  \ar@<0.5ex>[r]^{\omega_{1}}
& C_1 \ar@<0.5ex>[l]^{\omega_{1}} }
$ \, & \, 0.81 \, & \, 0.81 \, \\
\, $C_{1S}$ \, & \, -0.139 \, & 
\begin{minipage}{3.50cm} 
 \includegraphics[width=3.50cm]{FIG6.eps} 
\end{minipage}&
\, $\xymatrix{
C_{1S}  \ar@<0.5ex>[r]^{\omega_{2}}
& C_{1S} \ar@<0.5ex>[l]^{\omega_{2}} }
$ \, & \, 0.48 \, & \, 0.48 \,  \\
\, $C_2$ \, & \, 0.004 \, & 
\begin{minipage}{3.50cm} 
 \includegraphics[width=3.50cm]{FIG7.eps} 
\end{minipage}&
\, $\xymatrix{
C_1  \ar@<0.5ex>[r]^{\omega_{3}}
& C_2 \ar@<0.5ex>[l]^{\omega_{4}} }
$ \, & \, 0.61 \, & \, 0.47 \, \\
\, $C_3$ \, & \, 0.037 \, & 
\begin{minipage}{3.50cm} 
 \includegraphics[width=3.50cm]{FIG8.eps} 
\end{minipage}&
\, $\xymatrix{
C_1  \ar@<0.5ex>[r]^{\omega^{\prime}_{3}}
& C_3 \ar@<0.5ex>[l]^{\omega^{\prime}_{4}} }
$ \, & \, 0.65 \, & \, 0.48 \,  \\
\, $C_4$ \, & \, 0.019 \, & 
\begin{minipage}{3.50cm} 
 \includegraphics[width=3.50cm]{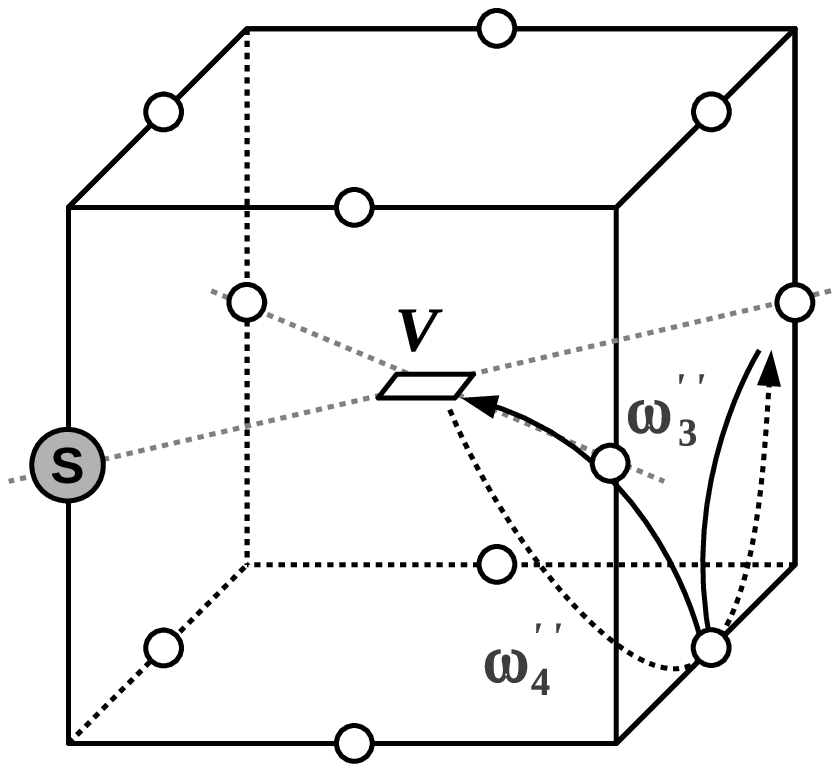} 
\end{minipage}&
\, $\xymatrix{
C_1  \ar@<0.5ex>[r]^{\omega^{\prime\prime}_{3}}
& C_4 \ar@<0.5ex>[l]^{\omega^{\prime\prime}_{4}} }$ \, & \, 0.73 \, & \, 0.58 \, \\
\, $C_5$ \, & \, 0.015 \, & 
\begin{minipage}{3.9cm} 
 \includegraphics[width=3.90cm]{FIG10.eps} 
\end{minipage}&
\, $\xymatrix{
C_2  \ar@<0.5ex>[r]^{\omega^{\star}_{0}}
& C_5 \ar@<0.5ex>[l]^{\omega^{\star}_{0}} }$ \, & \, 0.59 \, & \, 0.58 \, \\
\, $C_6$ \, & \, -0.003 \, & 
\begin{minipage}{3.90cm} 
 \includegraphics[width=3.90cm]{FIG11.eps} 
\end{minipage}&
\, $\xymatrix{
C_4  \ar@<0.5ex>[r]^{\omega^{\star}_{0}}
& C_6 \ar@<0.5ex>[l]^{\omega^{\star}_{0}} }$ \, & \, 0.63 \, & \, 0.65 \,  \\
\hline
\end{tabular}
\end{center}
\end{table}
\clearpage
For $Al-U$, as can be seen in Table \ref{T6}, the migration barriers are quite different from $0.65eV$, the value in perfect lattice, except for the transition $C_4 \rightarrow C_6$. In comparison with the $Ni-Al$ case, the jump $\xymatrix{C_1 \ar@<0.5ex>[r]^{\omega^{\prime\prime}_{3}} & C_4 \ar@<0.5ex>[l]^{\omega^{\prime\prime}_{4}} }$, involves more than one atom, as indicated in the figure inserted in Table \ref{T6}, and shown in more detail in Figure \ref{MJ}. In  Figure \ref{MJ}, we show both, direct and indirect jumps involving respectively one or two atoms. For the jump (1), the atom labeled 3 is dragged by the atom labeled 1 to the vacancy site. The jump (2) is the reverse of jump (1). While for the direct jump (3), the atom 1 jumps towards the vacancy, although, it is a high energy jump. 

As the direct jump (3) has lower probability of occurrence than the indirect jump (1), then present calculations of frequencies are performed using the values corresponding to this last one, that is   $0.73eV$ and $0.58eV$, to compute $\omega^{\prime \prime}_3$ and $\omega^{\prime \prime}_4$, respectively, and using $\nu^{\star}$ from Table \ref{Nu0}.

Although the jump $C_1 \rightarrow C_4$ in $Al-U$ involves two atoms it is not a successive jump. It is indeed a single jump which involves two atoms, that is, there is a single saddle point for the whole jump. The monomer method here employed is able to find both saddle point energy and configuration.

In table \ref{T7}, we show the migration barriers for more distant neighbors pairs than the forth. As can be seen, the values obtained are close to $0.65eV$, the migration barrier in the perfect crystal. 

\begin{table}[htdp] 
\begin{center}
\caption{Jumps beyond the second coordinated shell. The binding energies are shown in the second column. The third column denoted the frequency rate, where the superscripts $^{(\perp,\mp)}$ on $\omega_0$ implies vacancy jumps perpendicular $^{\perp}$, backward $^{-}$ or forward $^+$ in respect to the $\hat{x}$ direction. Migration energies are shown in column four and five.}
\label{T7}
\begin{tabular}{ccccc} 
\hline 
\, $C_n=S+V_n$ \, & \, $E_b(eV)$ \, & \, $\omega_i $ \, & \, $E^{\rightarrow}_m(eV)$ \, & \, $E^{\leftarrow}_m(eV)$ \, \\
\hline 
\hline
\, $C_7$ \, & \, 0.002 \, &  \, $\stackrel{\omega^{\perp}_{0}} {C_7\rightarrow C_{10}}$
\, & \, 0.61 \, & \, 0.64 \,  \\ 
\, $C_{8}$ \, & \, 0.015 \, & \, $\stackrel{\omega^{-}_{0}}{C_{8}\rightarrow C_{11}}$
\, & \, 0.64 \, & \, 0.61 \,  \\
\, $C_9$ \, & \, 0.002 \, & \, $\stackrel{\omega^{+}_{0}}{C_{12} \rightarrow  C_{12}} $ 
\, & \, 0.61 \, & \, 0.64 \, \\
\hline 
\end{tabular}
\end{center}
\end{table}

\begin{figure}[h]
\begin{center}
\includegraphics[width=8.0cm]{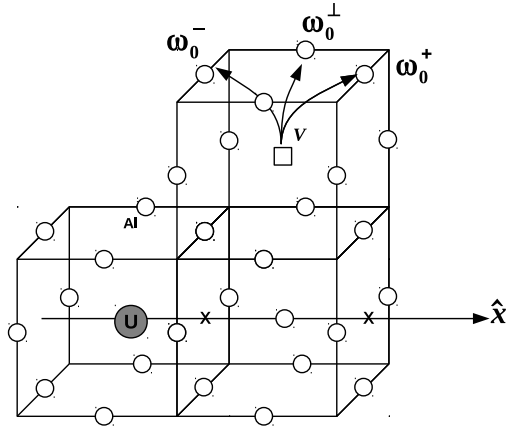}
\caption{Vacancy jumps beyond the second coordinated shell. The superscripts $^{(\perp,\mp)}$ on $\omega_0$ implies vacancy jumps perpendicular to, backward or forward $\mp \hat{x}$ respectively.}
\label{FIG14}
\end{center}
\end{figure}

In order to compute $\omega_i$, we use the conventional treatment formulated by Vineyard \cite{VIN57}, that corresponds to the classical limit, where the vibrational prefactors, $\nu^{\star}$, do not depend on the temperature, that is
\begin{equation}
\omega_{i}=\nu^{\star} \exp(-E^{\rightarrow}_m/k_BT),
\label{nu0T}
\end{equation}
with
\begin{equation}
\nu^{\star}=\frac{\displaystyle\prod_{i=1}^{3N} \nu^I_i }{\displaystyle\prod_{i=1}^{3N-1} \nu^S_i},
\label{PInu}
\end{equation}
and $E_m$ is the migration barrier. In (\ref{PInu}), $\nu^I_i$ and $\nu^S_i$ are the frequencies of the normal vibrational modes at the initial and saddle points, respectively. That is, $\nu^I_i$ refers to the vibrational frequencies of the nearest neighbors $X-V$ pair ($X$ = Ni, Al, U) and $\nu^S_i$ refers to the saddle configuration for the $S$-vacancy exchange, the product does not include the unstable mode.
Note that, Eq (\ref{PInu}) is based on calculation of the frequencies of the normal vibrational modes. This normal modes can involve only one atom or being collective modes. Hence it is also applicable to the single jump $C_1 \rightarrow C_4$ in $Al-U$ involving two atoms. \\

In Table \ref{Nu0} we report the calculated attempt frequencies. 

\begin{table}[htdp] 
\begin{center}
\caption{Attempt frequencies $\nu^{\star}$ in (\ref{nu0T}) in $THz$ unit. We compare present calculations with results using the density functional theory (DFT) respectively in the local density (LDA) and generalized gradient (GGA) approximations, and from Monte Carlo (MC) simulations.}
\label{Nu0}
\begin{tabular}{lcc|lcc} 
\hline 
Ref. & $Ni \rightarrow V$ in $Ni$ & $Al \rightarrow V$ in $Ni$ & Ref. & $Al \rightarrow V$ in $Al$ & $U \rightarrow V$ in $Al$ \\
\hline 
Present work & 23.7 & 30.8 & Present work & 19.56 & 8.25 \\
\cite{TUC10} DFT & 4.48 & - & \cite{MAN08} DFT (LDA) & 20.79 & - \\
\cite{DIV00} B2-$NiAl$ MC & 50.7  & 47.7 & \cite{MAN08} DFT (GGA) & 22.51 & -\\
  &  &  & \cite{SAN02} CMST & 22.60 & -\\
\hline 
\end{tabular}
\end{center}
\end{table}

Once the jump frequencies in the five-frequency model have been computed, the diffusion coefficients are calculated using analytical expressions in terms of the temperature. It is important to note the discrepancy between the classical and the quantum description concerning to the evaluation of $\omega_i$ \cite{TOY08}. Although these discrepancies are large in the low-temperature range the quantum value gradually converges to the classical one at temperatures higher than room temperature \cite{TOY08}. Hence, here we employ a classical description. 

Table \ref{T8} presents the calculated frequencies (\ref{nu0T}) for two different temperatures with the migration energies taken from Tables \ref{T5} and \ref{T6}. Using a different approach based on the Wert and Zener model \cite{WER65}, Zacherl \textit{et al.} \cite{ZAC12,TES12}, have studied diffusion in $Ni$ based diluted alloys using a temperature dependent frequency prefactor. 

\begin{table}[htdp]
\begin{center}
\caption{Vacancy jump frequencies $\omega_i$ calculated from (\ref{nu0T}) using the description of vineyard. The symbol ($^{\star}$) indicates effective frequencies.}
\label{T8}
\begin{tabular}{ccccc} 
\hline 
\, & \, $Ni-Al$ \, & & \, $Al-U$ \, &
\tabularnewline 
\, &\, $T_1=800K$\, & \, $T_2=1700K$\, & \, $T_1=300K$\, & \, $T_2=600K$\, 
\tabularnewline 
\hline 
\, $\omega_0$ \, & \, $1.6\times 10^{7}$ \, & \, $2.9\times 10^{10}$ \, & \, $2.9\times 10^{2}$ \, & \, $7.6\times 10^{7}$ \, 
\tabularnewline 
\, $\omega_1$ \, & \, $3.2\times 10^{6}$ \, & \, $1.4\times 10^{10}$ \, & \,\,\,\,\, $4.8\times 10^{-1}$ \, & \, $3.1\times 10^{6}$ \, 
\tabularnewline  
\, $\omega_2$ \, & \, $2.4\times 10^{7}$ \, & \, $4.1\times 10^{11}$ \, & \, $7.1\times 10^{4}$ \, & \, $7.7\times 10^{8}$ \, 
\tabularnewline 
\, $\omega^{\star}_3$ \, & \, $1.6\times 10^{7}$ \, & \, $2.9\times 10^{11}$ \, & \, $4.5\times 10^{2}$ \, & \, $8.3\times 10^{7}$ \, 
\tabularnewline 
\, $\omega^{\star}_4$ \, & \, $4.9\times 10^{8}$ \, & \, $5.0\times 10^{11}$ \, & \, $1.7\times 10^{5}$ \, & \, $1.7\times 10^{9}$ \, 
\tabularnewline 
\hline
\end{tabular}
\end{center}
\end{table}

From the calculated jump frequencies, then the tracer correlation factors $f_S$ and the solvent enhancement factors $b_{A^{\star}}$ can be obtained from (\ref{eq:FF}) and (\ref{SumRbA}), respectively.
They are shown in Table \ref{T9}, together with the jump frequencies ratios calculated according to the five-frequency model.

\begin{table}[h]
\begin{center}
\caption{Solvent enhancement and solute correlated factors for $Ni-Al$ and $Al-U$ at different temperatures. The first two columns describe the alloy and the temperature range considered. For the solvent enhancement factor $b_{A^{\star}}$ (column three), and for the solute correlated factor $f_S$ (column four). The last tree columns describe the jump frequency ratios of the solute$-$vacancy interaction.}
\label{T9}
\begin{tabular}{ccccccc}
\hline
\hline
\, Alloy \, & \, $T/K$ \, &  \, $b_{Ni^{\star}}$  \, & \, $f_{Al^{\star}}$ \, & \, $\omega_2/\omega_1$ \, & \, $\omega^{\star}_3/\omega_1$ \, & \, $\omega^{\star}_4/\omega_0$ \, \tabularnewline
\hline 
\, $Ni-Al$   \, & \, 700 \,  & \, -23.4 \, & \, 0.61 \, & \, 7.9 \, & \, 5.6 \, & \, 3.6 \, \tabularnewline 
\,  \,          & \, 800 \,  & \, -19.0 \, & \, 0.62 \, & \, 7.4 \, & \, 4.9 \, & \, 3.1 \, \tabularnewline 
\,  \,          & \, 900 \,  & \, -14.2 \, & \, 0.63 \, & \, 6.1 \, & \, 4.1 \, & \, 2.7 \, \tabularnewline 
\,  \,          & \, 1000 \, & \, -10.9 \, & \, 0.64 \, & \, 5.2 \, & \, 3.6 \, & \, 2.5 \, \tabularnewline 
\,  \,          & \, 1100 \, & \, -8.7 \, & \, 0.65 \, & \, 4.6 \, & \, 3.2 \, & \, 2.3 \, \tabularnewline 
\,  \,          & \, 1200 \, & \, -7.2 \, & \, 0.66 \, & \, 4.1 \, & \, 2.9 \, & \, 2.1 \, \tabularnewline 
\,  \,          & \, 1300 \, & \, -5.9 \, & \, 0.67 \, & \, 3.8 \, & \, 2.7 \, & \, 2.0 \, \tabularnewline 
\,  \,          & \, 1400 \, & \, -5.1 \,  & \, 0.67 \, & \, 3.5 \, & \, 2.5 \, & \, 1.9 \, \tabularnewline 
\,  \,          & \, 1500 \, & \, -4.4 \,  & \, 0.68 \, & \, 3.3 \, & \, 2.3 \, & \, 1.8 \, \tabularnewline 
\,  \,          & \, 1600 \, & \, -3.8 \,  & \, 0.68 \, & \, 3.1 \, & \, 2.2 \, & \, 1.8 \, \tabularnewline 
\,  \,          & \, 1700 \, & \, -3.3 \,  & \, 0.69 \, & \, 2.9 \, & \, 2.1 \, & \, 1.7 \, \tabularnewline 
\hline 
\hline
\, Alloy \, & \, $T/K$ \, & \, $b_{Al^{\star}}$  \, & \, $f_U^{\star}$ \, & \, $\omega_2/\omega_1$ \, & \, $\omega^{\star}_3/\omega_1$ \, & \, $\omega^{\star}_4/\omega_0$ \, \tabularnewline
\hline 
\, $Al-U$  \, & \, 300 \, & \, $-6.7\times 10^{3}$ \, & \, $6.4\times 10^{-3}$ \, & \, $147589.5$\, & \, $936.2$ \, & \, $151.5$ \, \tabularnewline 
\,  \, & \, 350 \, & \, $-2.9\times 10^{3}$ \, & \, $1.4\times 10^{-2}$ \, & \, $23826.5$ \,      & \, $333.8$ \, & \, $72.2$ \,  \tabularnewline 
\,  \, & \, 400 \, & \, $-1.6\times 10^{3}$ \, & \, $2.7\times 10^{-2}$ \, & \, $6068.2$ \,      & \, $155.3$ \, & \, $41.4$ \,\tabularnewline 
\,  \, & \, 450 \, & \, $-1.0\times 10^{3}$ \, & \, $4.5\times 10^{-2}$ \, & \, $2094.4$ \,      & \, $86.2$ \, & \, $26.9$ \, \tabularnewline 
\,  \, & \, 500 \, & \, $-6.7\times 10^{2}$ \, & \, $6.8\times 10^{-2}$ \, & \, $894.3$ \,                 & \, $53.9$ \, & \, $19.0$ \,\tabularnewline 
\,  \, & \, 550 \, & \, $-4.8\times 10^{2}$ \, & \, $9.6\times 10^{-2}$ \, & \, $445.7$ \,                 & \, $36.9$ \, & \, $14.4$ \, \tabularnewline 
\,  \, & \, 600 \, & \, $-3.6\times 10^{2}$ \, & \, $0.13$ \,              & \, $249.5$ \,                 & \, $26.9$ \, & \, $11.3$ \, \tabularnewline 
\,  \, & \, 650 \, & \, $-2.8\times 10^{2}$ \, & \, $0.16$ \,              & \, $152.7$ \,                 & \, $20.7$ \, & \, $9.3$ \, \tabularnewline 
\,  \, & \, 700 \, & \, $-2.2\times 10^{2}$ \, & \, $0.20$ \,              & \, $100.2$ \,                 & \, $16.5$ \, & \, $7.8$ \,  \tabularnewline 
\,  \, & \, 750 \, & \, $-1.8\times 10^{2}$ \, & \, $0.24$ \,              & \, $69.6$ \,                  & \, $13.6$ \, & \, $6.8$ \,  \tabularnewline  
\,  \, & \, 800 \, & \, $-1.4\times 10^{2}$ \, & \, $0.28$ \,              & \, $50.6$ \,                  & \, $11.5$ \, & \, $5.9$ \,  \tabularnewline  
\,  \, & \, 850 \, & \, $-1.2\times 10^{2}$ \, & \, $0.32$ \,              & \, $38.2$ \,                  & \, $9.9$ \, & \, $5.3$ \,  \tabularnewline 
\,  \, & \, 900 \, & \, $-1.0\times 10^{2}$ \, & \, $0.35$ \,              & \, $29.7$ \,                  & \, $8.9$ \, & \, $4.8$ \,  \tabularnewline 
\hline 
\end{tabular}
\end{center}
\end{table}

\begin{figure}[h]
\begin{center}
\includegraphics[angle=-90,width=12.0cm]{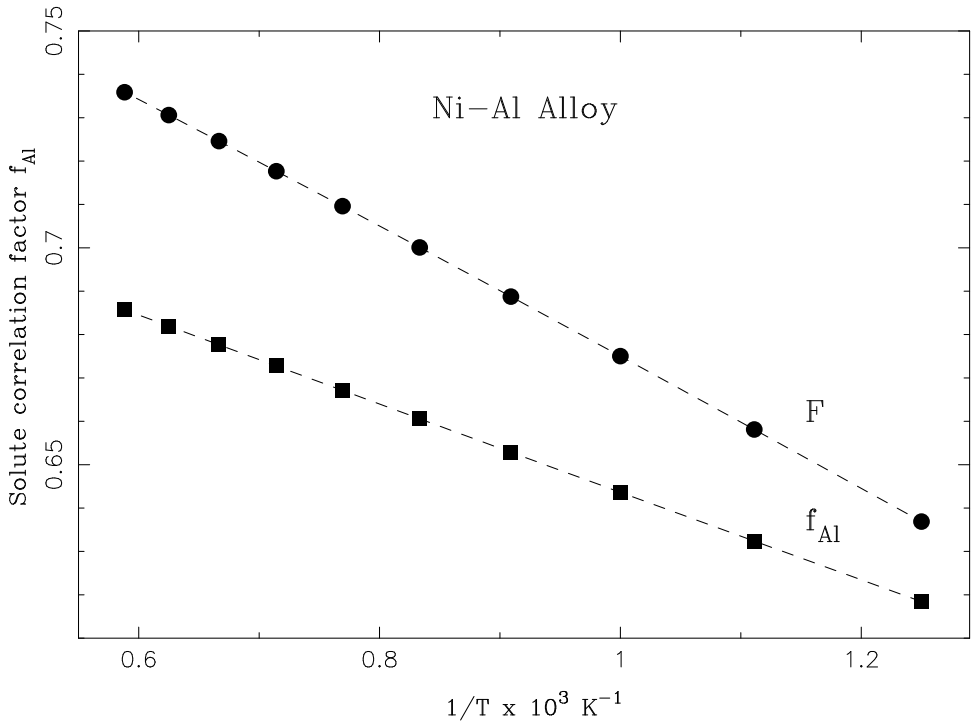}
\vspace{1.5cm}
\caption{Solute correlation factor $f_{Al^{\star}}$, obtained from (\ref{eq:FF}), in the $Ni-Al$ system as a function of the temperature in filled squares. The $F$ factor in (\ref{FMann}), is denoted with filled circles.}
\label{FIG15}
\end{center}
\end{figure}

\begin{figure}[h]
\begin{center}
\includegraphics[angle=-90,width=12.0cm]{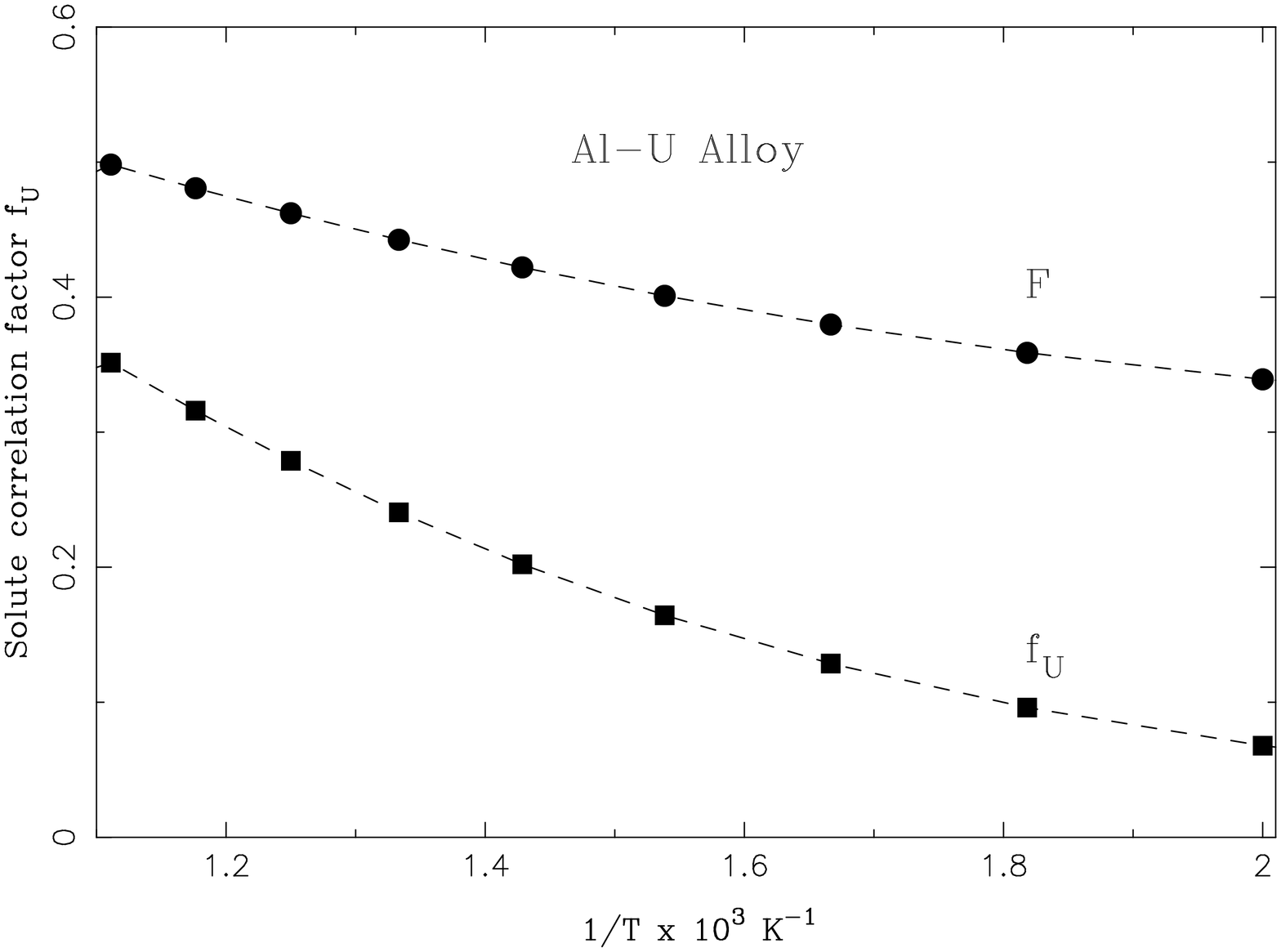}
\vspace{1.5cm}
\caption{Same as figure \ref{FIG15} for $Al-U$.}
\label{FIG16}
\end{center}
\end{figure}

The solute correlation factor, $f_S$, obtained from (\ref{eq:FF}), is also shown in Figures \ref{FIG15} and \ref{FIG16} in terms of the inverse of the absolute temperature, respectively for $Ni-Al$ and $Al-U$, together with the $F$ factor from (\ref{FMann}). 

\begin{figure}[h]
\begin{center}
\includegraphics[angle=-90,width=12.0cm]{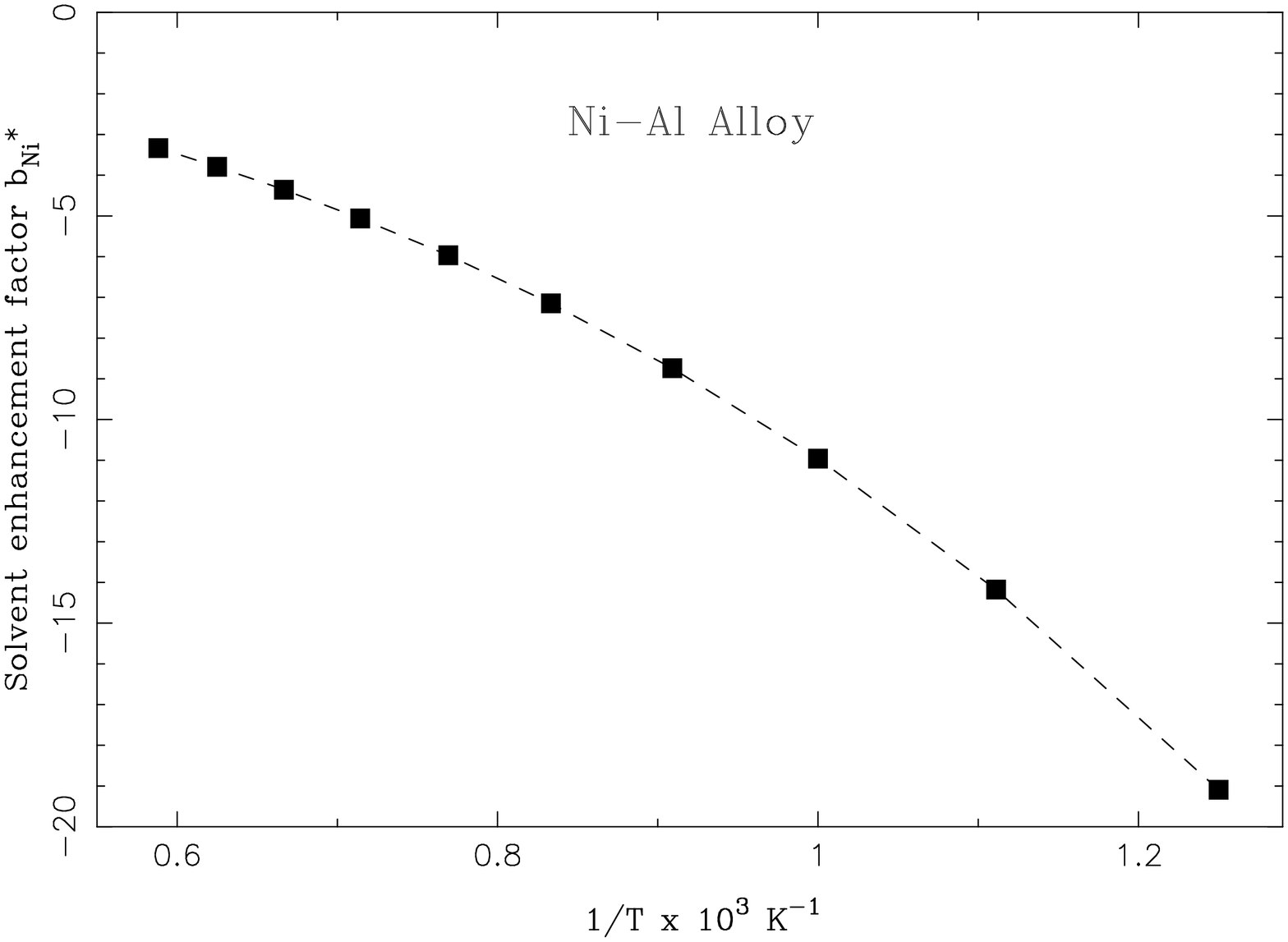}
\vspace{1.5cm}
\caption{Solvent-enhancement factor $b_{Ni}$ obtained from (\ref{SumRbA}), for the $Ni-Al$ system as a function of the temperature.}
\label{FIG17}
\end{center}
\end{figure}

\begin{figure}[h]
\begin{center}
\includegraphics[angle=-90,width=12.0cm]{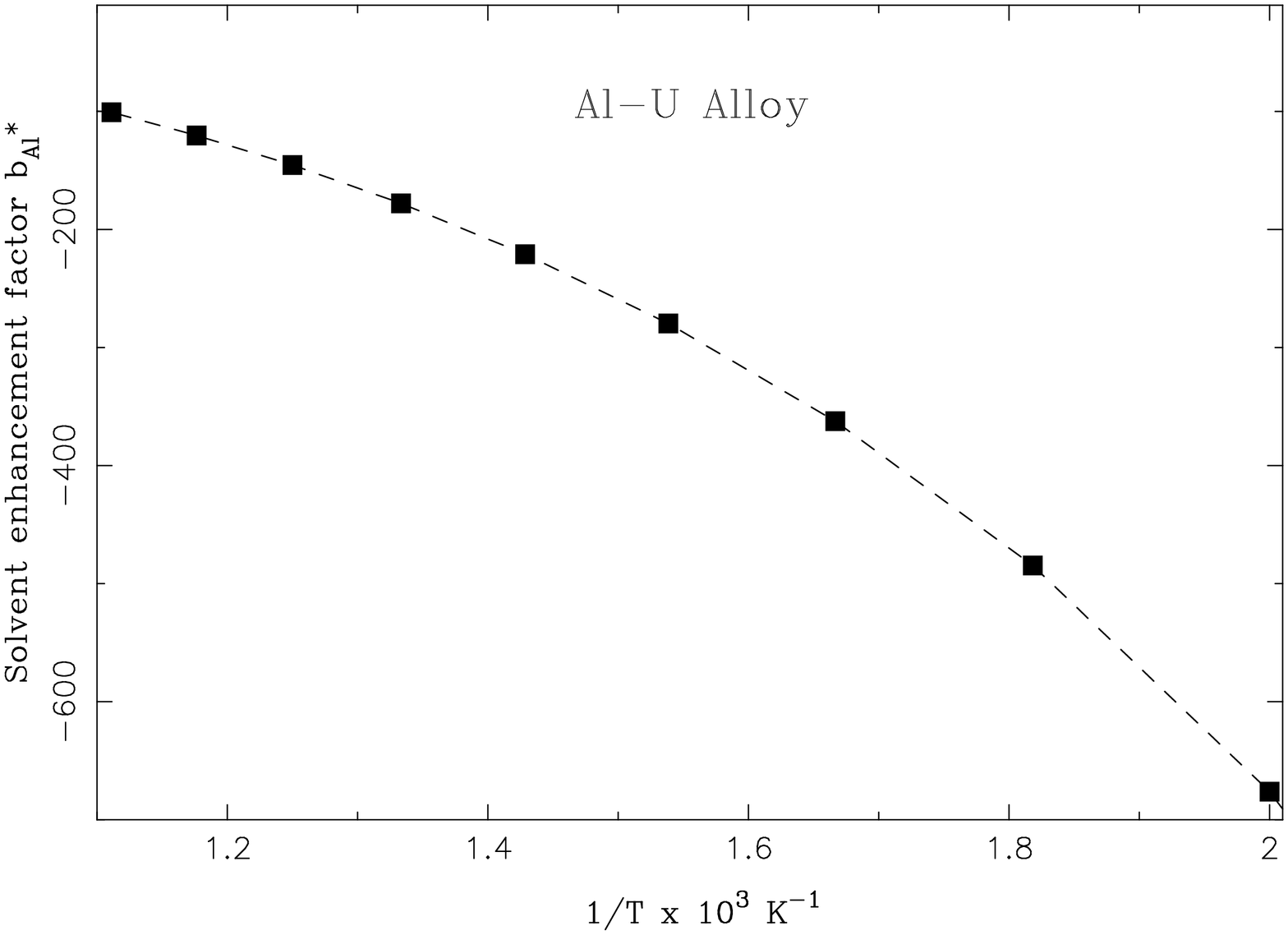}
\vspace{1.5cm}
\caption{Solvent-enhancement factor $b_{Al}$ obtained from (\ref{SumRbA}), for the $Al-U$ system as a function of the temperature.}
\label{FIG18}
\end{center}
\end{figure}

In Table \ref{T9}, the solvent-enhancement factors, $b_{A^{\star}}$, is obtained from (\ref{SumRbA}) and depicted in Figures \ref{FIG17}, \ref{FIG18}, respectively for $Ni-Al$ and $Al-U$, as a function of the temperature. It must be taken into account that the effect of $b_{A^{\star}}$ on the tracer self-diffusion coefficient $D^{\star}_A(c_S)$, must be multiplied by the solute concentration $c_S$, which is low for diluted alloys, hence $D^{\star}_A(c_S)$ is similar to $D^{\star}_A(0)$.


The Onsager and diffusion coefficients were calculated for a solute molar fraction $c_S=4.9\times 10^{-4}$, for both alloys, which corresponds to $n_{Al}=4.53\times 10^{19}cm^{-3} \, at./cm^3$ for $Ni-Al$ and $n_U=3.01\times 10^{19}cm^{-3} \, at./cm^3$ for $Al-U$. 

From $L_{AS}$ and $L_{SS}$, we also calculate the vacancy wind coefficient $G$ as in Ref. \cite{CHO11}. The $L_{VS}$ coefficient, which provides essential information about the flux of $S$ atoms induced by the vacancy flow can be defined in terms of the Onsager coefficients $L_{SS}$ and $L_{AS}$, respectively in (\ref{LBB1F}) and (\ref{LAB1F}) as,
\begin{equation}
L_{VS}= -(L_{SS}+L_{SA})=-L_{SS}(G+1),
\label{LBV}
\end{equation}
where $G$ is defined as the vacancy wind coefficient. The final expression is given by,
\begin{equation}
G= \frac{L_{AS}}{L_{SS}} = \frac{1}{(2\omega_1+7\omega^{\star}_3F)} \left[6\omega^{\star}_3-4\omega_1+14\omega^{\star}_3(1-F)\left( \frac{\omega_0-\omega^{\star}_4}{\omega^{\star}_4} \right) \right].
\label{GWind}
\end{equation}

\begin{figure}[h]
\begin{center}
\includegraphics[angle=-90,width=12.0cm]{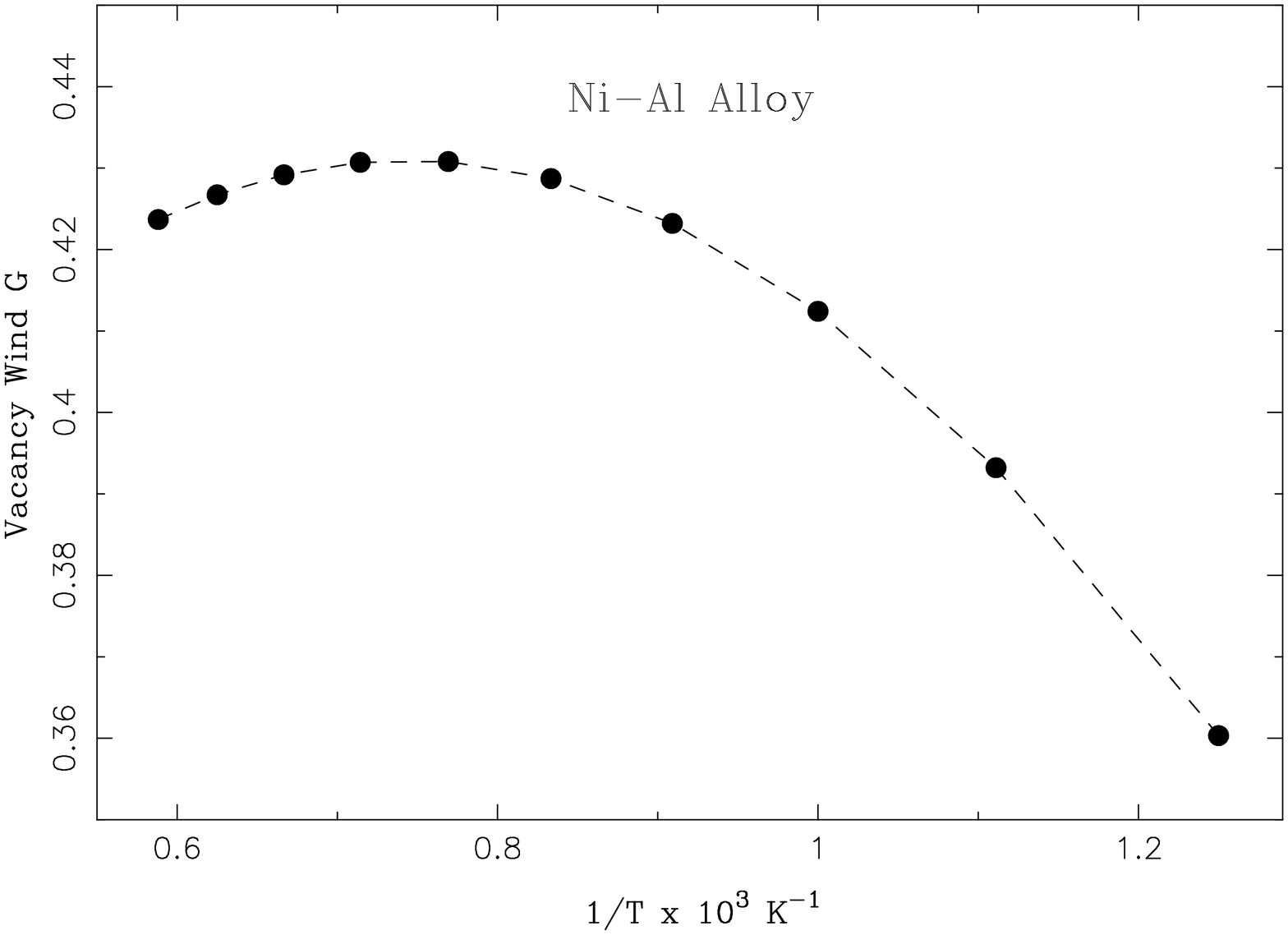}
\vspace{1.5cm}
\caption{The vacancy wind parameter $G$ in (\ref{GWind}): Ratio of the Onsager phenomenological coefficients of $Al$ in $Ni$ calculated from (\ref{LBB1F}) and (\ref{LAB1F}) \textit{vs} $1/T$.}
\label{FIG19}
\end{center}
\end{figure}

\begin{figure}[h]
\begin{center}
\includegraphics[angle=-90,width=12.0cm]{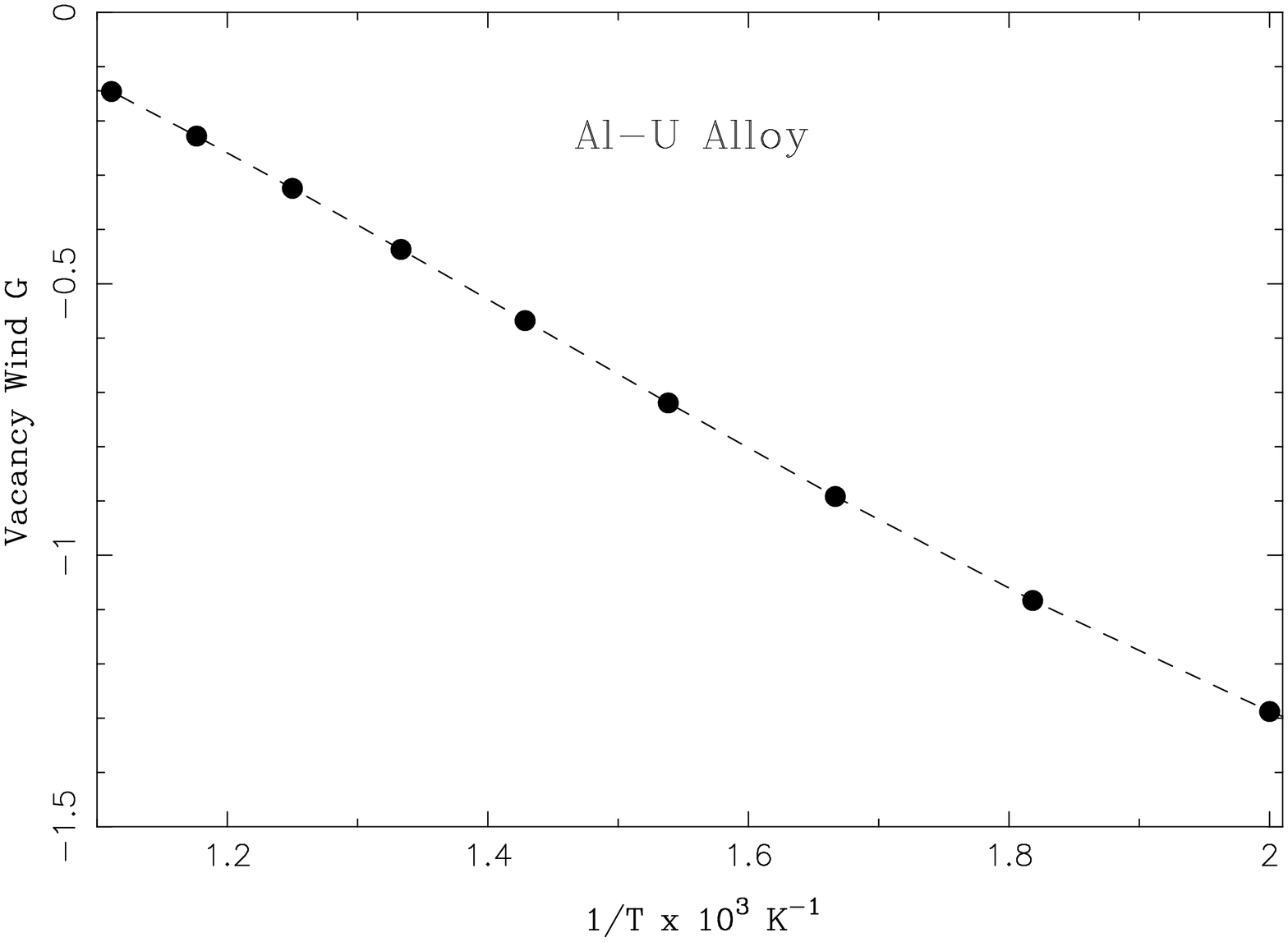}
\vspace{1.5cm}
\caption{The vacancy wind parameter $G$ in (\ref{GWind}): Ratio of the Onsager phenomenological coefficients of $U$ in $Al$ calculated from (\ref{LBB1F}) and (\ref{LAB1F}) \textit{vs} $1/T$.}
\label{FIG20}	
\end{center}
\end{figure}

The $G$ parameter in (\ref{GWind}) accounts for the coupling between the flux of species $J_A$ and $J_S$, through the vacancy flux, $J_V$ \cite{BOC96}. The results are presented in Figures \ref{FIG19} and \ref{FIG20}, for $Ni-Al$ and $Al-U$ systems respectively. In Figure \ref{FIG19}, the vacancy wind parameter verifies $G>-1$ for $Ni-Al$ in the full range of temperatures considered, while for $Al-U$, Figure \ref{FIG20} shows that $G>-1$ only above $550K$. 

In the case where $G<-1$, $L_{VS}$ is positive, then the vacancy and the solute diffuse in the same direction as a complex specie \cite{CHO11}. This transport phenomena could occur in $Al-U$ at lower temperatures, due to the strong binding of the $C_1$ pair, while is unlikely to occur for $Al$ in $Ni$ by the opposite argument.

\begin{figure}[h]
\begin{center}
\includegraphics[angle=-90,width=12.0cm]{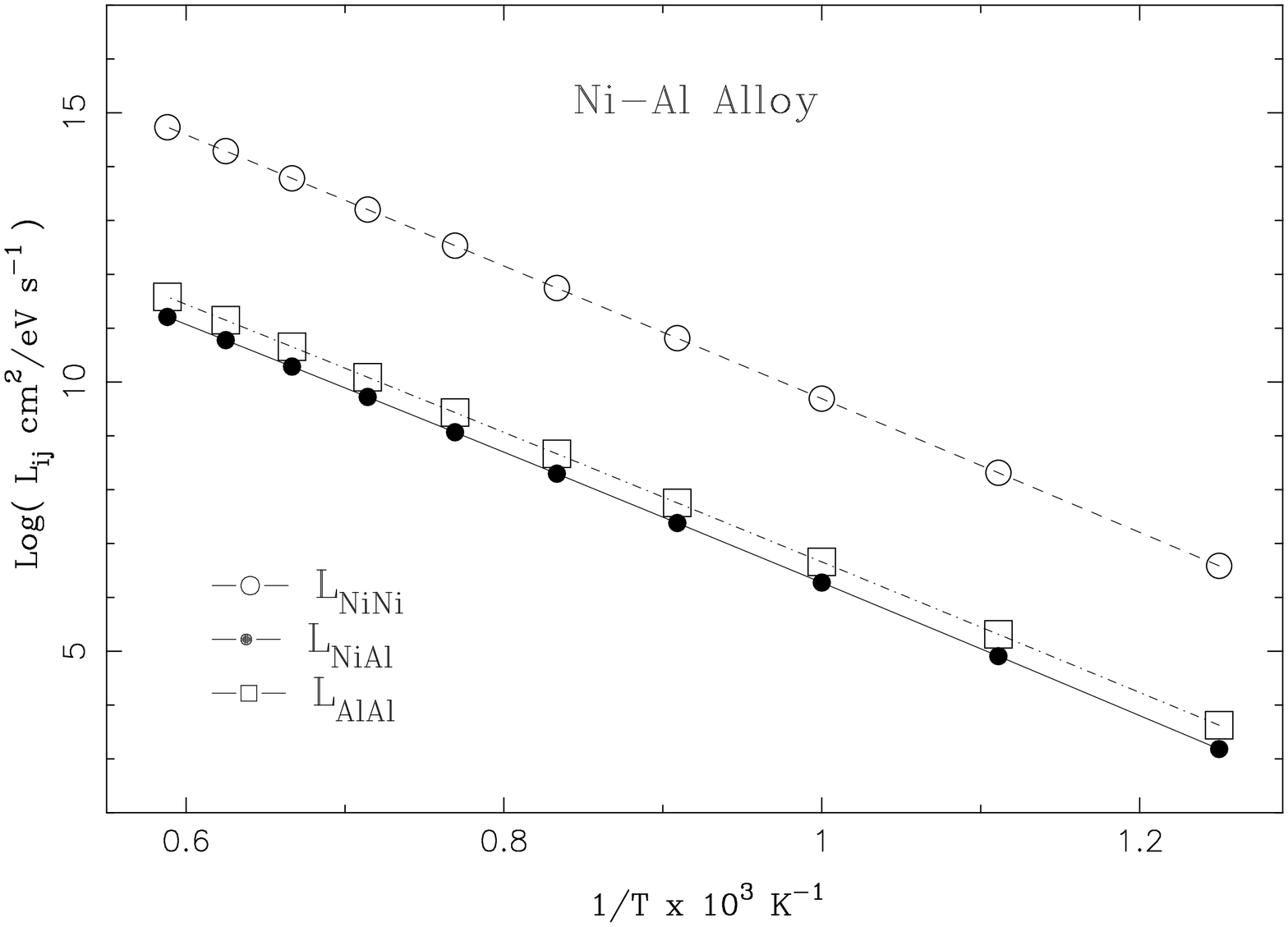}
\vspace{1.5cm}
\caption{Onsager phenomenological coefficients \textit{vs} $1/T$ for the $Ni-Al$ system. Squares denote $L_{AlAl}$, empty circles denote $L_{NiNi}$ while $L_{NiAl}$ is described with filled circles. The coefficients were calculated from (\ref{LBB1F}), (\ref{LAB1F}) and (\ref{LAA1F}).}
\label{FIG21}
\end{center}
\end{figure}

\begin{figure}[h]
\begin{center}
\includegraphics[angle=-90,width=12.0cm]{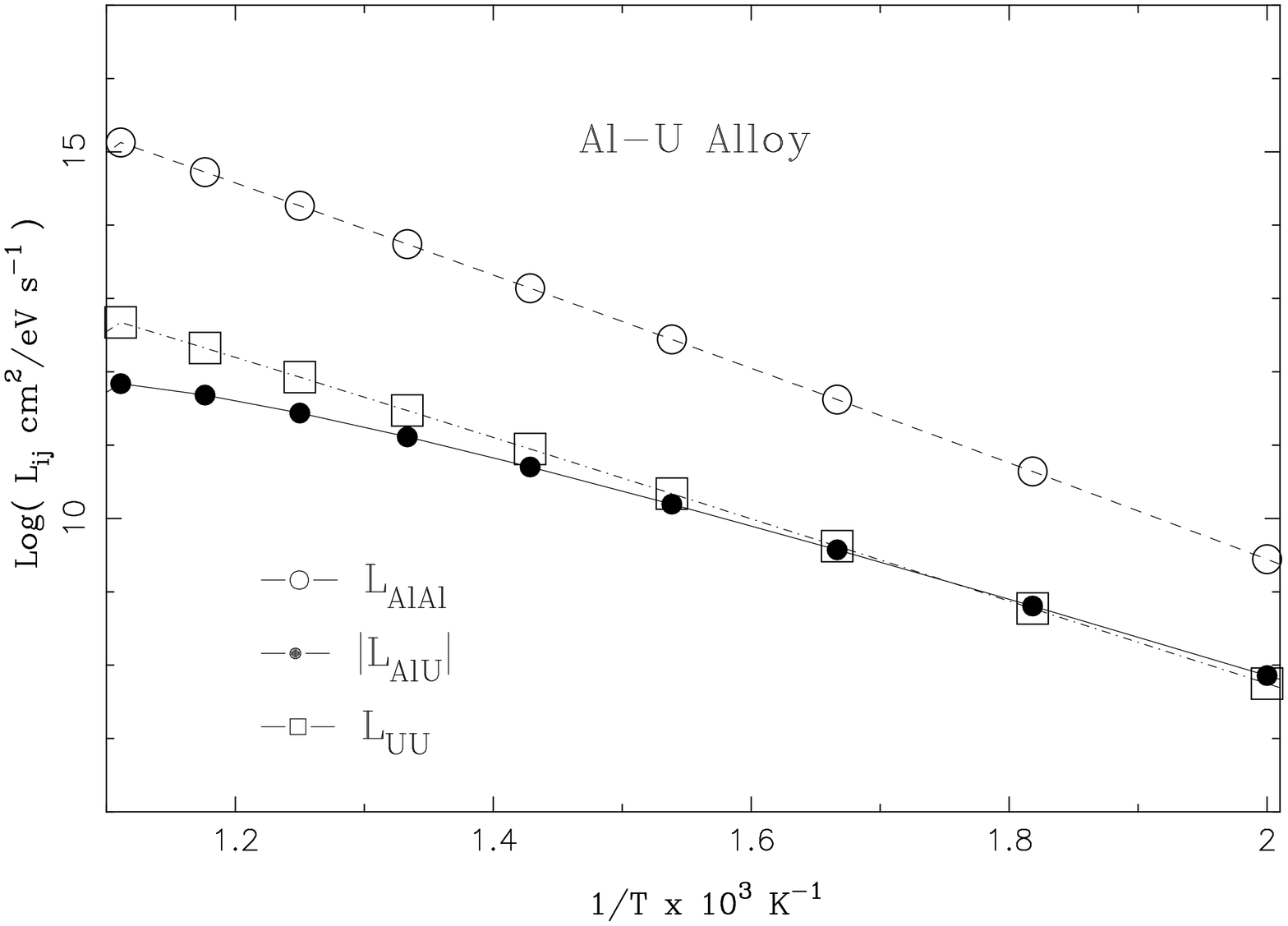}
\vspace{1.5cm}
\caption{Onsager phenomenological coefficients \textit{vs} $1/T$ for the $Al-U$ system. Squares denote $L_{UU}$, empty circles denote $L_{AlAl}$ while $|L_{UAl}|$ is described with filled circles. The coefficients were calculated from (\ref{LBB1F}), (\ref{LAB1F}) and (\ref{LAA1F}).}
\label{FIG22}
\end{center}
\end{figure}

The full set of $L$-coefficients, are displayed in Figs. \ref{FIG21} and \ref{FIG22}, against the inverse of the temperature for the $Ni-Al$ and $Al-U$, respectively. We see that for the $Ni-Al$ case the $L$-coefficients follow an Arrhenius behavior, which implies a linear relation between the logarithm of $L$-coefficients against the inverse of the temperature (see Fig. \ref{FIG21}). For $Al-U$ we can appreciate a deviation of the $L_{AlU}$ coefficient from the Arrhenius law at high temperatures (see Fig. \ref{FIG22}).   

In Figure \ref{FIG22}, the cross $L_{AlU}=L_{UAl}$ coefficient is negative in all the temperature range considered.

\begin{figure}[h]
\begin{center}
\includegraphics[angle=-90,width=12.0cm]{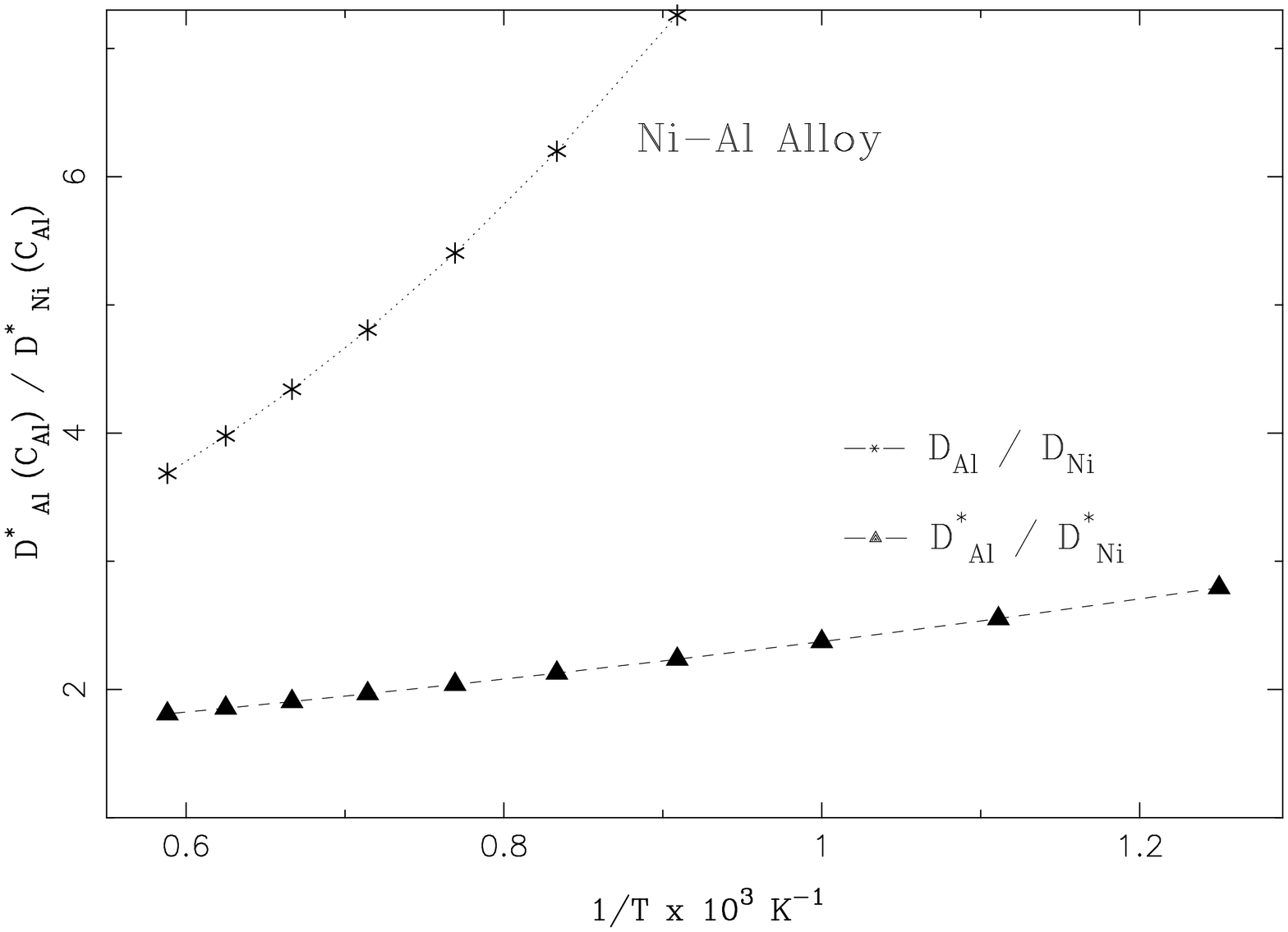}
\vspace{1.5cm}
\caption{Ratio of the tracer diffusion coefficient $D^{\star}_{Ni}/D^{\star}_{Al}$ in $Ni-Al$ \textit{vs} $1/T$. The ratio between the intrinsic diffusion coefficients, $D_{S}/D_{A}$ calculated from (\ref{DATEQ}) and (\ref{DBTEQ}), is also shown with symbols in asterisk and dashed line.}
\label{FIG23}
\end{center}
\end{figure}

\begin{figure}[h]
\begin{center}
\includegraphics[angle=-90,width=12.0cm]{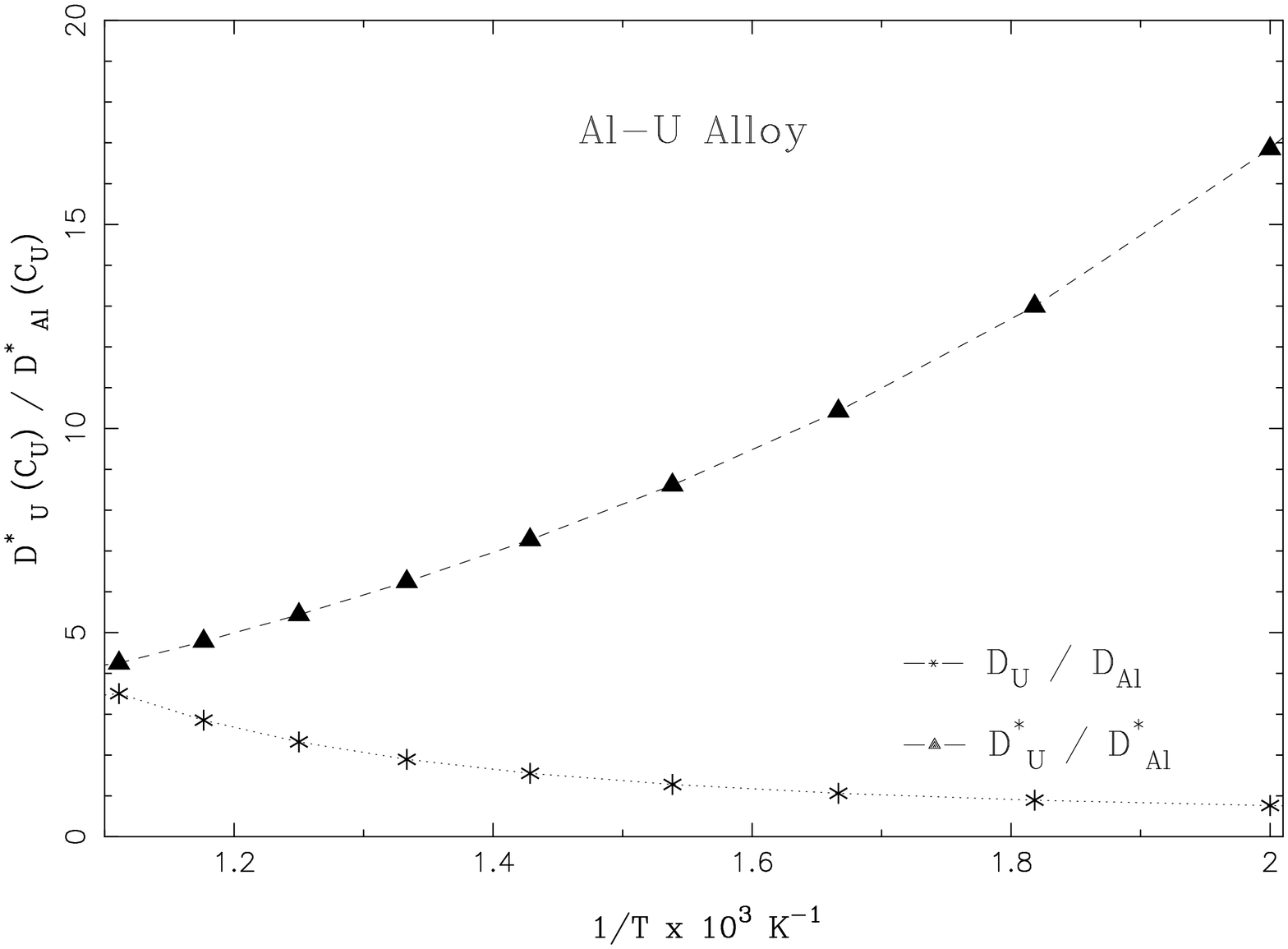}
\vspace{1.5cm}
\caption{Ratio of the tracer diffusion coefficient $D^{\star}_{U}/D^{\star}_{Al}$ in $Al-U$) \textit{vs} $1/T$. The ratio between the intrinsic diffusion coefficients, $D_{S}/D_{A}$ calculated from (\ref{DATEQ}) and (\ref{DBTEQ}), is also shown in stars.}
\label{FIG24}
\end{center}
\end{figure}

Now, we are in position to obtain the tracer diffusion coefficients $D^{\star}_A$ and $D^{\star}_B$. First, we present the ratio of the calculated tracer diffusion coefficients $D^{\star}_{S}/D^{\star}_{A}$ as a function of the inverse of the temperature for the $Ni-Al$ and $Al-U$ in Figures \ref{FIG23} and \ref{FIG24}, respectively. 

In Figures \ref{FIG23} and \ref{FIG24}, we also show the ratio between the intrinsic diffusion coefficients, $D_{S}/D_{A}$ (in stars symbols) calculated from (\ref{DATEQ}) and (\ref{DBTEQ}).

\begin{figure}[h]
\begin{center}
\includegraphics[angle=-90,width=12.0cm]{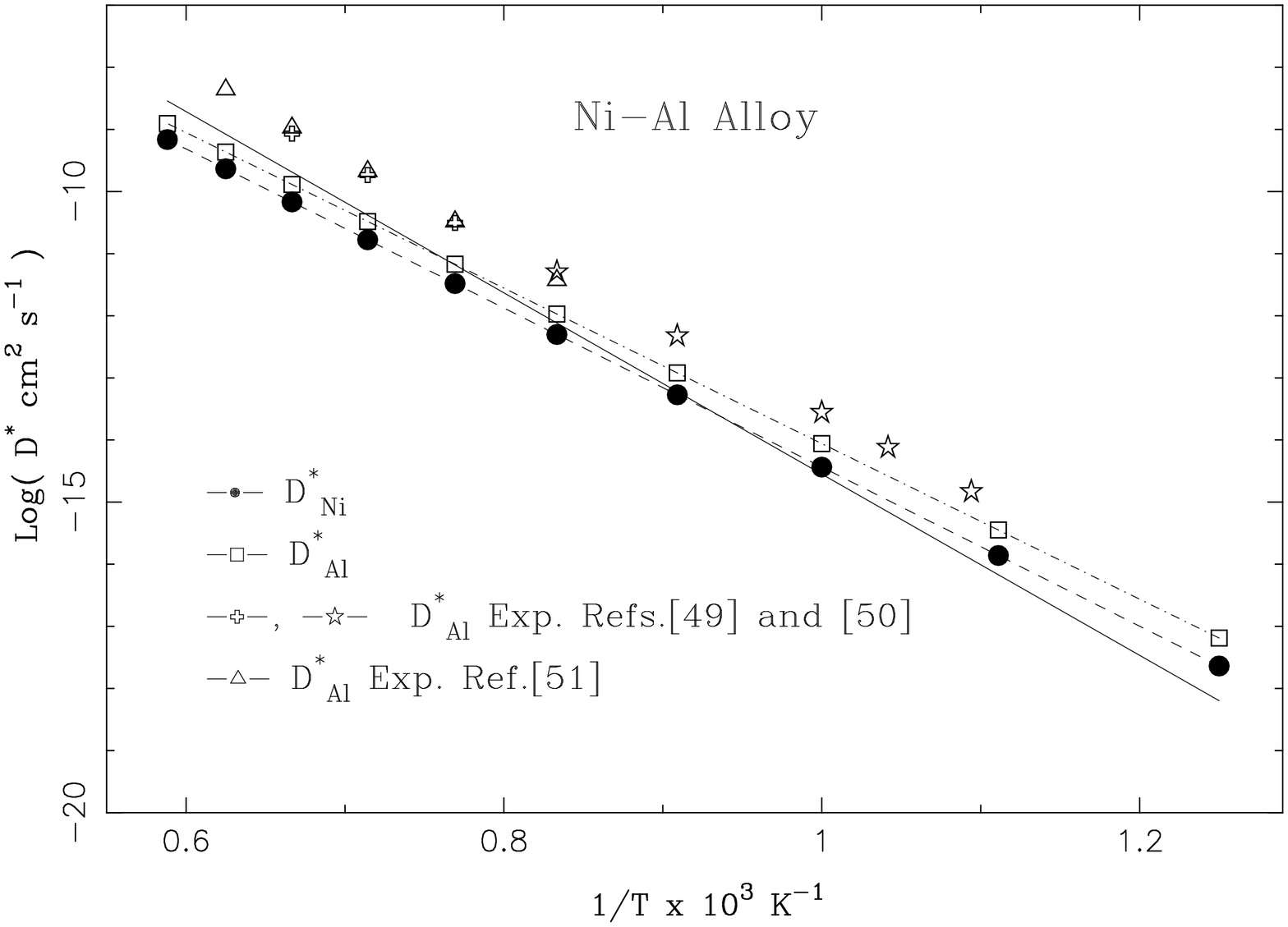}
\vspace{1.5cm}
\caption{Tracer diffusion coefficients of $Al$ ($D^{\star}_{Al}$ in open squares) and $Ni$ ($D^{\star}_{Ni}$ in filled circles) in the alloy, calculated from (\ref{DBB2}) and (\ref{DALC}), respectively. Solid line represents the best estimative of the pure $Ni$ self-diffusion coefficient $D^{\star,Exp}_{Ni}$, taken from Campbell work \cite{CAM11}. Available experimental data, for the $Al$ diffusion coefficient in the alloy, are displayed with stars \cite{GUS81} and cruxes \cite{SWA56}. In open triangles results from Ref. \cite{YAM80} for the solute tracer diffusion coefficient in a $\zeta -12\%$ mass $NiAl$ compound in the temperature range of $T=[1273-1573]^{\circ}C$.}
\label{FIG25}
\end{center}
\end{figure}

\begin{figure}[h]
\begin{center}
\includegraphics[angle=-90,width=12.cm]{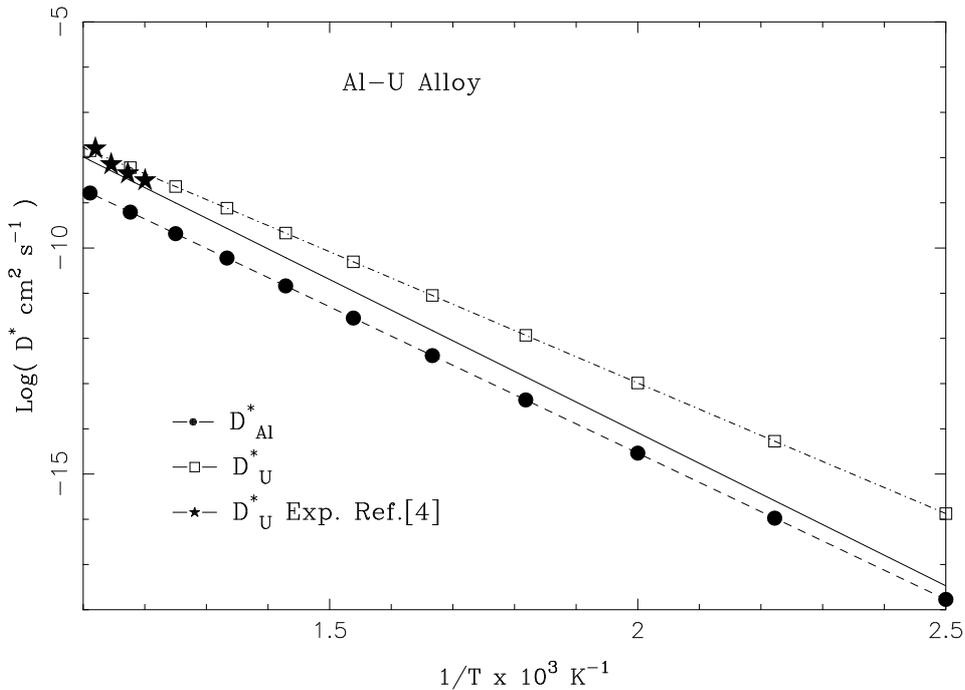}
\vspace{1.5cm}
\caption{Tracer diffusion coefficients of $U$ ($D^{\star}_U$ in open squares) and $Al$ ($D^{\star}_{Al}$ in filled circles) in the alloy, calculated from (\ref{DBB2}) and (\ref{DALC}), respectively. Solid line represents the best estimative of the pure $Al$ self-diffusion coefficient $D^{\star,Exp}_{Al}$, taken from Campbell work \cite{CAM11}. Available experimental data, for the $U$ diffusion coefficient in the alloy \cite{HOU71}, are displayed with filled stars.}
\label{FIG26}
\end{center}
\end{figure}

The tracer diffusion coefficients $D^{\star}_S$ and $D^{\star}_A$, calculated from (\ref{DBB2}) and (\ref{DALC}), are shown in Figures \ref{FIG25} and \ref{FIG26} respectively for $Ni-Al$ and $Al-U$. It is important to perform a comparison between theoretical results obtained in present work with reliable experimental data. We have verified that the tracer self diffusion coefficient $D^{\star}_A(c_S)$ for a diluted alloy is practically equal to that for the pure solvent $D^{\star}_A(0)$ (i.e., $D^{\star}_A(c_S) \simeq D^{\star}_A(0)$). 

Hence, we can test our results for $D^{\star}_A(c_S)$ with available experimental data in pure solvents. 

In this respect, Campbell et al. \cite{CAM11}, from a statistical analysis performed using weighted mean statistic, have determined a consensus estimators which best represents all known self diffusion available experimental data for pure solvent, $D^{\star,Exp}_{A}$. 
 
The estimator $D^{\star,Exp}_{A}$ corresponds to the experimental self-diffusivity of species $A$ in pure $A$ and is expressed in the form \cite{CAM11},
\begin{equation}
D^{\star,Exp}_{A}=D^0_A\exp(-Q_A/RT),
\label{DLCONS}
\end{equation}
where $R$ is the ideal gas constant, $T$ is the absolute temperature, while the values for $D^0_A$ and $Q_A$ in pure $Ni$ and $Al$, are taken from Ref. \cite{CAM11}, and are displayed in Table \ref{T10}.  

In order to perform a comparison of our results for $D^{\star}_A(c_S)$ with available experimental data in pure solvents, in Figures \ref{FIG25} and \ref{FIG26}, we display the calculated $Ni$ and $Al$ tracer self-diffusion coefficients (in filled circles and dashed lines), together with the consensus estimator $D^{\star,Exp}_{A}$ represented by solid lines. As can be observed, $D^{\star,Exp}_{A}$ fits well with the values of $D^{\star}_A$ calculated in the present work.

For the $Ni-Al$ system, Figure \ref{FIG25} also displays the tracer solute diffusion coefficient,  
our calculations (in open squares) are displayed together with experimental data for $T=[914-1212]^{\circ}C$ \cite{GUS81} and $T=[1372-1553]^{\circ}C$ \cite{SWA56} with stars and cruxes respectively. In open triangles, we also show the experimental results obtained by Yamamoto \textit{et al.} for inter-diffusion in a $\zeta -12\%$ mass $Al-Ni$ alloy in the temperature range of $T=[1273-1573]^{\circ}C$.

\begin{table}[htdp] 
\begin{center}
\caption{Parameters involved in the expression for the self-diffusion consensus fit $D^{\star,Exp}_{A}$, where the parameter $A$ indicates $Ni$ or $Al$ hosts. The first column denotes the reference where the values were taken from. The solvent lattice is indicated in the second column. The third and fourth columns denote the pre-exponential factor, $D^0_A$, and the activation energy, $Q_A$, for equation (\ref{DLCONS}) respectively. The range of temperatures of the description is referred in column five. The values were taken from Campbell work \cite{CAM11}.}
\label{T10}
\begin{tabular}{ccccc} 
\hline 
\, Ref. \, & \, Lattice \, & \, $D^0_A(cm^2s^{-1})$ \, & \, $Q_A(KJ/mol)$ \, & \, $T(^{\circ}C)$ \, \\
\hline 
\hline
\, \cite{CAM11} \, & \, $Ni$ \, & \, $1.1$ \, &  \, $279.35$ \, & \, $[769-1667]$ \,  \\ 
\, \cite{CAM11} \, & \, $Al$ \, & \, $0.292$ \, &  \, $129.7$ \, & \, $[357-833]$ \,  \\ 
\hline 
\end{tabular}
\end{center}
\end{table}

With respect to the $Al-U$ system, experimental values for the $U$ diffusion coefficient in $Al$ \cite{HOU71} at infinite dilution have been obtained by Housseau \textit{et al.} \cite{HOU71}. In Ref. \cite{HOU71}, the authors have obtained the diffusion parameters from the fit of their experimental permeation curves with the solution of the diffusion equation,
\begin{equation}
\frac{\partial C(x,t)}{\partial t}=D_U\frac{\partial ^2C(x,t)}{\partial x^2},
\label{FIT}
\end{equation}
with boundary condition $x=0;\, C(0,t)=S_0$, where $S_0$ is the maximum solubility of the diffusing specie in the alloy. They have proposed a solution for equation (\ref{FIT}) as,
\begin{equation}
C(x,t)=S_0[1-erf(x/2\sqrt{D_Ut}]. \label{C0U}
\end{equation}
Then the values of $D^{\star}_U$ and $S_0$ are obtained by fitting the experimental permeation curves with an expression of the form (\ref{C0U}).

The obtained diffusion parameters, taken from Ref. \cite{HOU71}, are shown in Table \ref{DEXP}, for different temperatures and $U$ concentrations, $c_U$. In their work \cite{HOU71}, the authors have concluded that, at infinite dilution, the dissolution of precipitates do not disturb the $U$ process diffusion in $Al$. 

\begin{table}[htdp] 
\begin{center}
\caption{Diffusion of $U$ in $Al$, for different temperatures ($1^{st}$ column) and $U$ molar concentrations $c_U$.}
\label{DEXP}
\begin{tabular}{ccccc} 
\hline 
\,  \, & \, \, & \, Uranium diffusion coefficient $D_U$&($\times10^{8}cm^2s^{-1}$) \, & \,  \,  \\
\, $T(^{\circ}C)$ \, & \, $c_U=2\times 10^{-3}$ \, & \, $c_U=9\times 10^{-4}$ \, & \, $c_U=2\times 10^{-4}$ \, & \, $c_U=6\times 10^{-7}$ \,  \\
\hline 
\hline
\, $620$ \, & \, $1.60 \pm 0.20$ \, &  \, $1.5 \pm 0.15$ \, & \, $1.56 \pm 0.15$ \, & \, $1.62 \pm 0.16$ \,  \\ 
\, $600$ \, & \, $0.78 \pm 0.08$ \, & \, $0.68 \pm 0.07$\, & \, $0.70 \pm 0.15$ \, & \, $0.65 \pm 0.07$ \,  \\
\, $580$ \, & \, $0.55 \pm 0.12$ \, & \, $0.70 \pm 0.12$ \, & \, $0.44 \pm 0.15$ \, & \, $0.67 \pm 0.10$ \, \\
\, $560$ \, & \, $0.40 \pm 0.10$ \, & \, $0.35 \pm 0.10$ \, & \, $0.31 \pm 0.10$ \, & \, $0.33 \pm 0.10$ \, \\
\hline 
\end{tabular}
\end{center}
\end{table}

In Figure \ref{FIG26}, we establish a comparison of our calculations for $D^{\star}_U$ with the experimental data in Table \ref{DEXP}, for a molar Uranium concentrations $c_U=2\times 10^{-4}$. We see that, experimental values (filled stars) in the temperature range of $[560-620]^{\circ}C$ are in perfect agreement with $D^{\star}_U$ obtained with the here described procedure. In the temperature range where there are available experimental data, the $U$ mobility is mainly due to direct interchange between the $U$ atom and the vacancy. 

On the other hand, the diffusion of $U$ in $Al$ was also calculated in a study of the maximum rate of penetration of $U$ into $Al$, in the temperature range $[473-663]^{\circ}C$ \cite{BIE55}. The maximum  penetration coefficient values in Ref. \cite{BIE55} were, $K_T=x^2/t=1.3\times 10^{-4}$, $8.8\times 10^{-5}$ and $1.1\times 10^{-8}$ $cm^2/s$ for $473^{\circ}C$, $523^{\circ}C$ and $663^{\circ}C$, respectively. From the expression $K=K_0\exp^{-Q/RT}$, the activation energy $Q$ was $Q=14.300$ in cal per mole in the temperature range of $[473-663]^{\circ}C$, where $R$ is expressed in calories per $1/^{\circ}C$ per mole, and $K_0$ is a proportionality constant. The plot $\ln K$ vs $1/T$ provides a convenient basis for expressing and comparing penetration coefficients. 

As a final comment, a recent work by Leenaers \textit{et al.} \cite{LEE08}, presents a great quantity of experimental findings for a real system, where the present model can also be applied. 

Also performed but not shown here, for the $Ni-Al$, we have reproduced all the microscopical parameters with $~100$ atoms using the classical molecular static technique and the SIESTA code coupled to the Monomer method \cite{RAM09}.

\clearpage

In the literature several researchers have studied the solvent atom-vacancy exchange in terms of the jump frequencies $\omega_i$ and $f_0$, in the framework of the random alloy model, as for example in Ref. \cite{BEL00}. The authors have performed an extensive Monte Carlo study of the tracer correlation factors in simple cubic, b.c.c. and f.c.c. binary random alloys. On the other hand, the kinetic formalism of Moleko \textit{et al.} \cite{MOL89}, also describes the behavior of the tracer correlation factors for slow and faster diffusers.


\section{Concluding remarks}
In summary, in this work we present the general mechanism based on non-equilibrium thermodynamics and the kinetic theory, to describe the diffusion behavior in f.c.c diluted alloys. 

Non equilibrium thermodynamic, through the flux equations, relates the diffusion coefficients with the Onsager tensor, while the Kinetic Theory relates the Onsager coefficients in terms of microscopical magnitudes. In this way we are able to write expressions for the diffusion coefficients only in terms of microscopic magnitudes, i.e. the jump frequencies.

The five frequency model has also been of great utility in order to discriminate the relevant jump frequencies, evaluated from the migration barriers under the harmonic approximation in the context of the conventional treatment by Vineyard corresponding to the classical limit. Hence, we have calculated the full set of phenomenological coefficients from which the full set of diffusion coefficients are obtained through the flux equation.

In this respect, the jump frequencies have been calculated from the migration barriers which are obtained with an economic static molecular techniques (CMST) namely the monomer method, that searches saddle configurations efficiently.

Although in this work we have performed the treatment for the case of f.c.c. latices where the diffusion is mediated by vacancy mechanism, a similar procedure can be adopted for other crystalline structures or different diffusion mechanism (for example, interstitials). 

We have exemplified our calculations for the particular cases of diluted $Ni-Al$ and $Al-U$ f.c.c. binary alloys. We have found that the tracer diffusion coefficient are in very good agreement with the available experimental data, for both alloys. 

Present calculations show that qualitatively a vacancy drag mechanism is unlikely to occur for the $Ni-Al$ system. In the case of $Al-U$, a vacancy drag mechanism could occur at temperatures below $550$K, while above this temperature the solute migrates by a direct interchange mechanism with the vacancy, such as was corroborated in the comparison with the available experimental data.

We have demonstrated that, the CMST is appropriate in order to describe the impurity diffusion behavior mediated by a vacancy mechanism in f.c.c. alloys. This opens the door for future works in the same direction where a similar procedure will be used that includes interstitial defects. 

\section*{Acknowledgments}

I am particularly grateful to Dr. Roberto C. Pasianot for help on calculations of the attempt jump frequencies, to Dr. A.M.F. Rivas for comments on the manuscript, and to Mart\'in Urtubey for Figure \ref{FIG2}. This work was partially financed by CONICET PIP-00965/2010.

\end{document}